\documentclass[12pt,preprintnumbers,nofootinbib,letterpaper]{article}
\pdfoutput=1
\usepackage{amsmath,amssymb,amsfonts,cite,xcolor,epsf,epsfig,epstopdf,graphics,graphicx,mathtools,subcaption,physics,verbatim}
\def\thefootnote{*\arabic{footnote}}
\definecolor{ultramarine}{rgb}{0.07, 0.04, 0.56}
\definecolor{cadmiumgreen}{rgb}{0.0, 0.42, 0.24}
\definecolor{indigo(dye)}{rgb}{0.0, 0.25, 0.42}
\usepackage[linktocpage=true,breaklinks]{hyperref}
\hypersetup{
	colorlinks=true,
	citecolor=ultramarine,
	linkcolor=cadmiumgreen,
	urlcolor=indigo(dye),
}

\usepackage[utf8]{inputenc}

\numberwithin{equation}{section}

\usepackage{array}
\newcolumntype{P}[1]{>{\centering\arraybackslash}p{#1}}

\usepackage[shortlabels]{enumitem}

\usepackage{dcolumn}
\newcolumntype{M}[1]{>{\centering\arraybackslash}m{#1}}
\newcolumntype{N}{@{}m{0pt}@{}}

\usepackage{amsmath}
\usepackage{ulem}

\newcommand{\Mpl}{M_{\rm Pl}}
\newcommand{\cs}{c_s}
\newcommand{\m}{m}

\newcommand{\D}{{\rm d}}

\newcommand{\be}{\begin{equation}}  
\newcommand{\ee}{\end{equation}}

\usepackage{tikz}
\begin{document}

\begin{flushright} {\footnotesize YITP-24-178}  \end{flushright}

\begin{center}

\def\thefootnote{\fnsymbol{footnote}}

\vspace*{1.5cm}
{\Large {\bf Dark matter from inflationary quantum fluctuations}}
\\[1cm]

{Mohammad Ali Gorji$^{1}$, Misao Sasaki$^{2,3,4,5}$, Teruaki Suyama$^{6}$}
\\[.7cm]

{\small \textit{$^1$Cosmology, Gravity, and Astroparticle Physics Group, Center for Theoretical Physics of the Universe, Institute for Basic Science (IBS), Daejeon, 34126, Korea
}}\\

{\small \textit{$^{2}$Kavli Institute for the Physics and Mathematics of the Universe (WPI), The University of Tokyo, \\ Chiba 277-8583, Japan}}\\

{\small \textit{$^{3}$Asia Pacific Center for Theoretical Physics (APCTP), Pohang 37673, Korea}}\\

{\small \textit{$^{4}$ Center for Gravitational Physics and Quantum Information, Yukawa Institute for Theoretical Physics, \\ Kyoto University, Kyoto 606-8502, Japan}}\\

{\small \textit{$^{5}$Leung Center for Cosmology and Particle Astrophysics, National Taiwan University, Taipei 10617, Taiwan}}\\

{\small \textit{$^{6}$Department of Physics, Institute of Science Tokyo, 2-12-1 Ookayama, Meguro-ku, \\
Tokyo 152-8551, Japan}}\\

\end{center}

\vspace{1.5cm}

\hrule \vspace{0.5cm}

\begin{abstract}
We explore a scenario in which dark matter is a massive bosonic field, arising solely from quantum fluctuations generated during inflation. In this framework, dark matter exhibits primordial isocurvature perturbations with an amplitude of ${\cal O}(1)$ at small scales that are beyond the reach of current observations such as those from the CMB and large-scale structure. We derive an exact transfer function for the dark matter field perturbations during the radiation dominated era. Based on this result, we also derive approximate expressions of the transfer function in some limiting cases where we confirm that the exact transfer function reproduces known behaviors. Assuming a monochromatic initial power spectrum, we use the transfer function to identify the viable parameter space defined by the dark matter mass and the length scale of perturbations. A key prediction of this scenario is copious formation of subsolar mass dark matter halos at high redshifts. Observational confirmation of a large population of such low-mass halos will support for the hypothesis that dark matter originated purely from inflationary quantum fluctuations.
\end{abstract}
\vspace{0.5cm} 

\hrule
\def\thefootnote{\arabic{footnote}}
\setcounter{footnote}{0}

\thispagestyle{empty}

\newpage
\hrule
\tableofcontents
\addtocontents{toc}{\protect\setcounter{tocdepth}{2}} 
\vspace{0.7cm}
\hrule

\newpage

\section{Introduction}\label{introduction}
The standard $\Lambda$CDM model of cosmology provides a comprehensive explanation for cosmological and astrophysical observations. 
According to this model, the matter content of the Universe consists of the cosmological constant, dark matter, and ordinary matter that includes baryonic matter and radiation. 
The cosmological constant $\Lambda$ is more generically called dark energy by allowing its time variation. The existence of dark energy is necessary to explain the observed late time cosmic acceleration \cite{SupernovaSearchTeam:1998fmf,SupernovaCosmologyProject:1998vns}. 
The current understanding is that dark energy constitutes about 70\% of the energy content of the Universe. 
The dark matter, which is needed to explain the observed rotation curves of the galaxies as well as the structure formation in the Universe~\cite{Liddle:1993fq,Bertone:2004pz,Feng:2010gw,Ferreira:2020fam}, contributes about 27\%, while the visible sector, consisting of the ordinary baryonic matter and radiation, contributes only 3\% of the energy content of the Universe \cite{Planck:2018vyg}. 
Although the $\Lambda$CDM model is observationally successful, it has its own shortcomings from the theoretical point of view. 
We have no compelling theoretical explanation for the current value of the cosmological constant $\Lambda$ or the dark energy density, which is called the cosmological constant problem \cite{Weinberg:1988cp,Peebles:2002gy}.
A more subtle mystery is the origin of dark matter, which remains unknown.

The scenario of cosmic inflation, in which the Universe undergoes a phase of quasi-exponential expansion at its very early stage, successfully explains the origin of all the observed structures in the Universe. The quantum vacuum fluctuations of a scalar field, which is typically the one that drives inflation, called the inflaton, are stretched to scales much greater than the Hubble radius during inflation and provide the seed for the density perturbation of the Universe~\cite{Mukhanov:2005sc,Weinberg:2008zzc}.

As dark energy and dark matter constitute around 97\% of the content of the Universe, 
it is very tempting to consider a scenario in which the dark energy and/or dark matter are also produced during inflation. 
However, this is not an easy task. Thus, most of scenarios deal with the production of the dark energy and dark matter particles after inflation, i.e. during the radiation dominance. 
Nevertheless, there has been some efforts to construct scenarios for the production of dark energy \cite{Barausse:2005nf,Kolb:2005me,Hirata:2005ei,Ringeval:2010hf,Glavan:2015cut,Glavan:2017jye} and dark matter particles \cite{Polarski:1994rz,Chung:1998zb,Chung:2001cb,Chung:2004nh} during inflation.

Recently, there have been some attempts to consider the production of dark matter from superhorizon perturbations of spectator fields during inflation \cite{Graham:2015rva,Kolb:2017jvz,Ema:2018ucl,Alonso-Alvarez:2018tus,Li:2019ves,Alonso-Alvarez:2019ixv,Cembranos:2019qlm,Ling:2021zlj,Firouzjahi:2021lov,Kolb:2022eyn,Garcia:2023qab,Racco:2024aac}. 
Depending on the spin, mass, and interactions between the spectator fields and inflationary sector (inflaton and gravity), different dark matter scenarios are suggested. 
In the case of spin-0 dark matter, the production mechanism is very similar to that of the curvature perturbation on superhorizon scales \cite{Graham:2015rva,Kolb:2017jvz,Ema:2018ucl,Alonso-Alvarez:2018tus,Li:2019ves,Alonso-Alvarez:2019ixv,Cembranos:2019qlm,Ling:2021zlj,Firouzjahi:2021lov,Kolb:2022eyn,Garcia:2023qab,Racco:2024aac},
while the production of dark matter particles with higher spins is more subtle. 
The reason is that the approximate isometries of the quasi-de Sitter inflationary background may prevent the superhorizon production of the light (with mass smaller than the Hubble parameter during inflation) higher spin particles. 
However, assuming a sizable interaction between a spectator field and the inflaton, one can overcome this issue and construct higher spin dark matter, like spin-1 (vector) dark matter \cite{Bastero-Gil:2018uel,Ema:2019yrd,Alonso-Alvarez:2019ixv,Nakai:2020cfw,Ahmed:2020fhc,Kolb:2020fwh,Salehian:2020asa,Firouzjahi:2020whk,Gross:2020zam,Ling:2021zlj,Bastero-Gil:2021wsf,Redi:2022zkt,Sato:2022jya,Bastero-Gil:2022fme,Nakai:2022dni,Bastero-Gil:2023htv,Bastero-Gil:2023mxm,Ozsoy:2023gnl,Cembranos:2023qph,Capanelli:2024nkf} and spin-2 dark matter \cite{Kolb:2023dzp,Gorji:2023cmz,Kolb:2023ydq}. 

To link the initial conditions of the spectator field set during inflation with late-time observables such as the present-day abundance, it is essential to understand how the field fluctuations evolve after inflation. Assuming that the modes of our interest re-enter the horizon during radiation dominance, one has to derive the transfer function for them from the time when they are on superhorizon scales until the time when they are well within the horizon.
While gravitational growth is suppressed during this period, the evolution of these modes is already non-trivial.
This complexity arises from three distinct incidents that take place during the radiation dominated era: 
(1) the modes transition from a relativistic to a non-relativistic regime, 
(2) they re-enter the Hubble horizon, and (3) the field starts oscillating due to its mass term.
The chronological order of these events depends on the interplay between the mass of the field, wavelength of the modes, and the Hubble parameter.
Although some of previous studies provide analytic insights into the time evolution of the field during the radiation dominated era, they are based on the approximations under which the different regimes are treated separately in a rather ad-hoc manner.
To the best of our knowledge, an analytic transfer function of a massive field during radiation dominance, given in terms of the exact solution of the field equation, has not been presented in the literature.
One of the primary goals of this paper is to derive an exact transfer function that automatically encodes all the aforementioned regimes in a unified manner.
The result has multiple applications.
For instance, it enables us to estimate the mass ranges relevant to dark matter scenarios quantitatively.
It can also be used as the initial condition for numerical simulations of the late-time structure formation.

In this paper, we consider a phenomenological scenario which covers many dark matter scenarios with different spins and interactions, and compute the relic density of the dark matter today. 
In the next two sections, we describe our setup and explain how various shapes of the initial power spectrum can arise during inflation.
In Sec.~\ref{sec:transfer-func}, we present an exact analytic transfer function and derive approximate expressions for several limiting cases.
Sec.~\ref{sec:energy-density} provides the $\Omega$ parameter of the dark matter field in various regimes,
based on the obtained transfer function.
In Sec.~\ref{sec:dark-matter}, we combine these results with
various observational constraints on dark matter to determine the allowed region of the parameter space 
of our phenomenological model.
The final section provides a summary. Some technical parts are presented in the appendices.


\section{The model}\label{sec-model} 

We tacitly assume that inflation is governed by a single scalar field. But other than that, we adopt an agnostic view about the dynamics of the inflaton and its coupling to the spectator field $X(t,{\bf x})$. 
Since the spectator field $X$ remains subdominant during inflation, we ignore the metric perturbation and 
account only for its couplings to the inflaton $\phi$.
Thus we treat $X$ as a quantum field in the expanding background.
We also ignore the self-interactions of $X$ and higher-derivative terms.
Then the action we consider is given by
\begin{align}\label{action}
S = \frac{1}{2} \int \D^3{ x}\,\D{t}\, a^3 f^2
\left[
\dot{X}^2 - \frac{\cs^2}{a^2} \left(\partial_iX\right)^2 -\m^2 X^2
\right] \,,
\end{align}
where a dot denotes the derivative with respect to the cosmic time $t$, $a(t)$ is the scale factor, $\cs$ is the sound speed, $\m$ is the mass, and $f(t)$ is a function of time that represents an interaction with the inflaton. 
In general, $\cs$ and $\m$ can also depend on time, though we assume that they are constant for the sake of simplicity. 
The explicit functional form of the dimensionless coupling $f(t)$ depends on the model under consideration.

We are interested in quantum fluctuations of the field $X$, assuming that the vacuum expectation value (vev) of $X$ is negligible. 
In cosmological background, the modes with wavelengths larger than the Hubble horizon can be treated as a homogeneous vev and one naturally expects $\langle{ {\hat X}}\rangle\neq0$ in our observable universe.\footnote{ 
Here we have placed $~{\hat {}}~$ to make it clear that the field is an operator.
The state for which the expectation values are computed is assumed to be the one annihilated 
by the annihilation operators.}
Thus the proper statement of our assumption will be
\begin{align}\label{X-vev}
\langle {\hat X}^2 \rangle \gg \langle {\hat X} \rangle^2 
\quad
\Rightarrow
\quad
\sigma_X^2 
= \langle {\hat X}^2 \rangle - \langle {\hat X} \rangle^2
\approx \langle {\hat X}^2 \rangle \,,
\end{align}
where $\sigma_X^2$ is the variance of $X$. 
In practice, we can simply consider $\langle{\hat X}\rangle=0$ keeping in mind that by $\langle{\hat X}\rangle=0$ we mean that $\langle{\hat X}\rangle$ is negligible compared to $\sqrt{\langle {\hat X}^2 \rangle}$. 
Working with conformal time $\tau=\int\D{t}/a(t)$ and going to Fourier space 
\begin{align}\label{Fourier}
{\hat X}(\tau,{\bf x}) =  \int \frac{\D^3{ k}}{{(2\pi)}^{3/2}}~\exp(i{\bf k}\cdot {\bf x}) 
{\hat X}_{\bf k}(\tau),
\end{align}
we expand the Fourier amplitude as ${\hat X}_{\bf k}(\tau) = X_k(\tau) \, \hat{a}_{\bf k} + X^{\ast}_k(\tau) \, \hat{a}_{-\bf k}^{\dagger}$ where  $X_k(\tau)$ is the mode function, $\hat{{a}}_{\bf k}$ and $\hat{{a}}_{\bf k}^{\dagger}$ are annihilation and creation operators which satisfy the usual commutation relations $[\hat{{a}}_{\bf k},\hat{{a}}_{\bf q}^{\dagger}] = \delta({\bf k}-{\bf q})$. 
We then find
the mode function satisfies
\begin{align}\label{EoM}
X''_k + 2 \frac{\left(af\right)'}{af}\, X'_k + \left( \cs^2k^2+\m^2a^2\right) X_k &= 0 \,,
\\
X_kX^*{}'_k-X^*_kX'_k&=\frac{i}{f^2a^2}\,.
\end{align}

Although the original action \eqref{action} in the configuration space deals with a spin-0 scalar field, the corresponding action for the mode functions in the Fourier space effectively includes spin-1 (vector) and spin-2 cases. 
More precisely, the Fourier space action can be identified with each helicity of spin-1 and spin-2 fields and our results in this paper can then be applied accordingly. For example, in the case of a minimally coupled scalar field, we have $f=1$ \cite{Firouzjahi:2021lov}, while for the transverse modes of a vector field, $f=a^{-1}$ \cite{Nakai:2020cfw,Firouzjahi:2020whk} (see also \cite{Gorji:2023cmz} for the spin-2 case). 
For the moment, we keep $f$ to be a general function of time.

The dimensionless power spectrum, defined as usual $\langle{\hat X}_{\bf k} {\hat X}_{\bf q} \rangle = \left(2\pi^2/k^3\right) {\cal P}_X(k,\tau)  \delta({\bf k}+{\bf q})$, is given by
\begin{align}\label{PS-def}
{\cal P}_X(k,\tau)=\frac{k^3}{2\pi^2} \big| X_{k} \big|^2 \,.
\end{align}
The energy density of the field $X$ is given by
\begin{align}\label{rho-x}
{\hat \rho}_X(\tau,x) = \frac{f^2}{2a^2} \left[
{\hat X}'^2 + \cs^2 \left(\partial_i {\hat X} \right)^2 +\m^2 a^2 {\hat X}^2
\right] \,.
\end{align}
The above expression is obtained from the Hamiltonian density constructed out of the quadratic action \eqref{action}. It is worth mentioning that, for a given full action including gravity, the energy density is defined by the energy-momentum tensor. These two definitions for the energy density may not always coincide (see for instance \cite{Firouzjahi:2021lov,Gorji:2023cmz}). For a model with given full action including gravity, one can always compute the corresponding energy-momentum tensor and find the energy density accordingly. If that energy density coincides with the one constructed out of the Hamiltonian density, all of our results in this paper can be applied. Otherwise, one needs to follow our strategy with the new expression for the energy density.

Going to the Fourier space \eqref{Fourier} and taking vacuum average of \eqref{rho-x}, we find the homogeneous averaged energy density 
\begin{align}\label{rho}
\bar{\rho}_X(\tau) = \frac{f^2}{2a^2} \int \frac{\D^3k}{(2\pi)^3}\, 
\left[ 
\big| X'_{k} \big|^2 + \left( \cs^2 k^2 + \m^2 a^2 \right) \big| X_{k} \big|^2
\right] \,.
\end{align}
One can then define spectral dimensionless energy density $\Omega_X(\tau,k)$ as
\begin{align}\label{Omega-def}
\Omega_X(k,\tau) \equiv
\frac{1}{3\Mpl^2H^2} \frac{\D{\bar \rho}_X}{\D\ln{k}}
=
\frac{f^2k^3}{12\pi^2\Mpl^2a^2H^2} 
\left[ 
\big| X'_{k} \big|^2 + \left( \cs^2 k^2 + \m^2 a^2 \right) \big| X_{k} \big|^2
\right] \,.
\end{align}
It is also useful to define the dimensionless vacuum average energy density as
\begin{align}\label{Omega-homogeneous}
{\bar \Omega}_X(\tau) = \int \Omega_X(\tau,k) \D\ln{k} = \frac{{\bar \rho}_X}{3\Mpl^2H^2}\,,
\end{align}
where we assume the existence of a UV cutoff $k<k_{\rm max}$ in the integral, which will be determined in the next section. 

\section{Superhorizon modes during inflation}
\label{sec:superhorizon}

In our scenario, the initial condition is set during inflation. 
In order to make it possible for the quantum fluctuations of $X$ to be frozen on superhorizon scales, we need to assume that $X$ is a light field during inflation,\footnote{One may notice that assuming \eqref{massless-inf} makes it difficult to achieve \eqref{X-vev}. This is a general feature of models based on dark matter production from frozen superhorizon perturbations. Therefore, a mechanism is required to reconcile these conditions. We propose possible scenarios in Sec.~\ref{summary}.}
\begin{align}\label{massless-inf}
\m\ll{H}_{\rm inf} \,,
\end{align}
where $H_{\rm inf}$ is the Hubble parameter during inflation which we assume to be constant for simplicity. 
Applying this condition to Eq. \eqref{EoM}, we find
\begin{align}\label{EoM-inf}
&X''_{k} + 2 \frac{\left(af\right)'}{af}\, X'_{k}  + \cs^2k^2 X_{k} = 0 \,.
\end{align}
In flat spacetime, where both $a$ and $f$ are constant, the second friction-like term in the l.h.s. disappears and the above equation has oscillating solution of the form $e^{\pm{i}\cs{k}\tau}$ which characterize the usual vacuum fluctuations. On the other hand, in the time-dependent FLRW background, the friction term can make the modes freeze. 
In the long-wavelength limit $c_s k\to 0$, two independent solutions are
$X_k ={\rm const.}$ and the time-dependent part of $X_k =\int^\tau d\tau'/(a^2 f^2)$ with the latter
being a quickly decaying function. 

Thus only the constant mode remains at late times.
Practically, the modes freezes once the friction term dominates over the gradient term in \eqref{EoM-inf}. Thus, we consider the modes that satisfy
\begin{align}\label{tachyonic-modes}
\cs{k} < \frac{\left(af\right)'}{af} \,,
\end{align}
for some period during inflation. This condition gives an upper limit $k_{\max}$ on $k$, and
any modes with $k \lesssim k_{\rm max}$ acquire perturbations which eventually freeze.
The explicit value of $k_{\max}$ can be determined for a given functional form of $f$.
For instance, if $f=1$, the condition \eqref{tachyonic-modes} is satisfied only for the superhorizon modes $\cs{k}\ll{a}H_{\rm inf}$ or $-\cs{k}\tau\ll1$ where we have used $a\simeq-1/(H_{\rm inf}\tau)$. Here we mention that by horizon we mean the sound horizon $c_sk=a'/a$.
If $f=a^{-1}$, which happens for spin-1 fields, then it is not possible to satisfy \eqref{tachyonic-modes}. This is a well-known fact that spin-1 fields will not feel the de Sitter horizon due to the conformal symmetry. In that case, having $f\neq a^{-1}$ (non-minimal coupling which can be achieved, i.e., through 
direct interactions with the inflaton field) is necessary to break the conformal symmetry and generate an effective horizon for vector modes \cite{Watanabe:2009ct,Nakai:2020cfw,Firouzjahi:2020whk}. 
Thus, for $f\neq a^{-1}$, condition \eqref{tachyonic-modes} can be satisfied for the modes which are not necessarily superhorizon.

As a scale-invariant power spectrum for $X$ shows up when $f=1$, the power spectrum of $X$ can be strongly scale-dependent if $f$ is time-dependent. 
For example, using the WKB approximation for the modes deep inside the horizon, $f^2$ appears in the denominator of the corresponding power spectrum. 
If we assume that $f=1$ everywhere except at the time $\tau=\tau_p$ where $f$ has a dip, there appears a peak in the power spectrum at $k_p=-1/(\cs\tau_p)$ \cite{Pi:2021dft}. 
Indeed, as we will show below, this simple form of $f$ is enough for our purpose.

After specifying the form of $f$ for a given model, we can solve Eq. \eqref{EoM-inf} and find the power spectrum \eqref{PS-def} during inflation. 
We assume that $f=1$ after inflation.\footnote{For the canonical vector field,
$f=a^{-1}$. Even for this case, the solution of the wave equation in the radiation dominated era is obtained by suitably rescaling the solution for $f=1$.} 
Then, the power spectrum of superhorizon modes $X$ at the onset of the radiation dominated era $\tau_i$ can be obtained from \eqref{PS-def} as
\begin{align}\label{PS-inf}
{\cal P}_{X,i} (k) 
\equiv {\cal P}_X(k,\tau_i) 
= \frac{k^3}{2\pi^2} \big| X_{k,i} \big|^2 \,,
\end{align}
where $X_{k,i}\equiv{X}_{k}(\tau_i)$ denotes the value of $X_{k}(\tau)$ at superhorizon regime $\cs k \tau_i\to0$.\footnote{To be more precise, if we match the scale factor at the end of inflation to the radiation stage at $\tau=\tau_i (<0)$, the conformal time for the radiation-dominated universe will be slightly shifted as $a\propto (\tau-2\tau_i)$. Nevertheless, for the modes of our interest $-c_sk\tau_i\ll1$, we may ignore this shift and assume $a\propto \tau$.} 
In the rest of this paper, we call (\ref{PS-inf}) initial power spectrum. The whole role of the coupling function $f$ is to have different shapes and scale-dependency for the initial power spectrum 
imprinted during inflation.

\section{Transfer function: Evolution after inflation}
\label{sec:transfer-func}

The purpose of this section is to provide a general analytic expression of the transfer function of $X$ in the radiation dominated era as a solution to (\ref{EoM}). The transfer function obtained in this section will be used to compute the energy density of $X$,  defined in Eq. \eqref{Omega-def},
in Sec.~\ref{sec:energy-density}. The readers who are only interested in the application of our setup to dark matter may directly move to Sec.~\ref{sec:dark-matter}.

We assume that $X$ is a subdominant component during the radiation domination and, therefore, the corrections to the metric perturbations are suppressed until the time of the matter-radiation equality.
Thus, we treat $X$ as a test field on the FLRW background and do not include 
the metric perturbations sourced by the fluctuations of $X$ as well as
the adiabatic perturbations originating from the inflaton fluctuations.

We define the transfer function of $X$ as 
\begin{align}\label{transfer-function-def}
T(k,\tau) = \frac{X_{k}(\tau)}{X_{k,i}} \,.
\end{align}
Then, (\ref{EoM}) yields the differential equation of the transfer function;
\begin{align}\label{Transfer-function}
T''(k,\tau) + 2 \frac{a'}{a} T'(k,\tau) + \left( \cs^2k^2+\m^2a^2\right) T(k,\tau) = 0 \,,
\end{align}
where, as we have mentioned previously, 
we have set $f=1$ during the radiation dominance. The solution for the transfer function with scale factor $a=a_i(\tau/\tau_i)$ is given by
\begin{align}\label{t-transfer-0}
T_m(k,\tau) = e^{-\frac{i\m}{2H}} \, _1F_1\left(\frac{3}{4}+\frac{iH_k}{4 \m };\frac{3}{2}; \frac{i\m}{H} \right)
 \,,
\end{align}
where the subscript $m$ indicates the mass dependence, $_1F_1$ is Kummer's function (the confluent hypergeometric function of the first kind) and
\begin{align}\label{H-i}
H_k \equiv \frac{\cs{k}}{a_k} \,,
\end{align}
is the Hubble parameter evaluated at the time when the mode with $k$ re-enters the sound horizon during the radiation dominance, $\cs{k}=a_kH_k=1/\tau_k$. The integration constants are fixed such that $T_m(k\tau\to0)=1$. Precisely speaking, the number of relativistic degrees of freedom may vary in time.
However, for simplicity, we ignore this time variation as it is a minor effect. 

In general, we have three dimensionful parameters in our setup: The Hubble parameter $H$, the mass $m$, and the scale (momentum) of our interest $k$ or the corresponding Hubble parameter at horizon re-entry $H_k$.\footnote{Here and below, by horizon we mean the sound horizon.} 
However, the transfer function \eqref{t-transfer} turns out to be a function of only two dimensionless variables $m/H$ and $m/H_k$ which are constructed out of $H$, $k$, $m$. 
These two dimensionless variables can be represented in a unified manner by defining the following new variables:
\begin{align}\label{mu-def}
\mu \equiv \frac{m}{H} \,,
\qquad
\mu_k \equiv \frac{m}{H_k} \,.
\end{align}
The transfer function \eqref{t-transfer-0} then can be rewritten as
\begin{align}\label{t-transfer}
T_m(\mu,\mu_k) = e^{-\frac{i\mu}{2}} \, _1F_1\left(\frac{3}{4}+\frac{i}{4 \mu_k };\frac{3}{2}; i\mu \right)
\,,
\end{align}
such that $\mu$ controls the time evolution and $\mu_k$ determines the scale dependence of the transfer function. We also define another dimensionless time variable,
\begin{align}\label{x-def}
x \equiv \sqrt{\frac{\mu}{\mu_k}} = \sqrt{\frac{H_k}{H}} = \cs{k}\tau \,,
\end{align}
which is not independent of $\mu$ and $\mu_k$, but turns out to be very useful. In particular, $x>1$ and $x<1$ correspond to the subhorizon and superhorizon modes respectively. Note that $x$ is independent of the mass.

To the best of our knowledge, this paper is the first to derive the analytic transfer function given by (\ref{t-transfer}) based on the exact solution of (\ref{EoM}). 
This is one of the main results of the paper. 
Although (\ref{t-transfer}) is exact, its form itself does not give us an intuitive understanding of its behavior. 
In the following subsections, we derive approximate but much simpler expressions for the transfer function in some limiting cases. 
By doing so, we also confirm that the exact transfer function reproduces the known qualitative behaviors.

\subsection{Relativistic regime}
The relativistic modes are those that have physical momenta much larger than the mass. More precisely, they satisfy
\begin{align}\label{modes-R}
&\m \ll \frac{\cs{k}}{a} \,,
\qquad
\Rightarrow 
\qquad
\mu=\frac{m}{H} \ll \frac{H_k}{m}=\mu_k^{-1}  \,.
\end{align}
In obtaining the second inequality, we have used $m^2 \ll c_s^2 k^2/a^2 = a_k^2 H_k^2/a^2=HH_k$.
Thus, the relativistic condition \eqref{modes-R} is rewritten as
\begin{align}\label{modes-R-mu}
\mu\,\mu_k \ll 1 \,.
\end{align}
For $\mu_k\ll1$ and $\mu\ll1$, this condition is trivially satisfied. 
However, when $\mu>1$ or $\mu_k>1$,
the relativistic condition \eqref{modes-R-mu} can only be satisfied 
if $\mu_k$ is sufficiently small (for $\mu > 1$) or $\mu$ is sufficiently
small (for $\mu_k >1$), respectively, or vice versa.
Thus, we first look at the two different limits $\mu_k\ll1$ and $\mu\ll1$. 

For $\mu_k\ll1$ with a fixed value of $\mu$ such that the relativistic condition \eqref{modes-R-mu} is satisfied, the first argument in the function ${}_1F_1(a;b;z)$ is large, $|a|\gg1$. 
As shown in appendix \ref{app-Kummer}, this leads to the expression given in Eq. \eqref{Hypergeometric-Larg}. 
In our case, setting $b=3/2$, we find
\begin{align}\label{Slater-TF}
T_m(k,\tau)
=
\frac{\Gamma\left(1-a\right)}{\Gamma\left(\frac{3}{2} - a\right)}
\left[ \frac{\sin \left(2 i a^{\frac{1}{2}} z^{\frac{1}{2}}\right)}{2z^{\frac{1}{2}}} \sum_{s=0}^{\infty}\frac{p_{s}(z)}{a^{s}} - \frac{i}{2a^{\frac{1}{2}}} \left(\cos \left(2 i a^{\frac{1}{2}} z^{\frac{1}{2}}\right)
-\frac{\sin\left(2i a^{\frac{1}{2}} z^{\frac{1}{2}}\right)}{2 i a^{\frac{1}{2}} z^{\frac{1}{2}}}\right) \sum_{s=0}^{\infty}\frac{q_{s}(z)}{a^{s}}
\right]
,
\end{align}
where 
\begin{align}\label{a-z-def}
a \equiv \frac{i}{4\mu_k} + \frac{3}{4} \,,
\qquad
z \equiv i\mu \,,
\end{align}
and $p_s(z)$ and $q_s(z)$ are defined in \eqref{p-def} and \eqref{q-def}. 
Note that the series form \eqref{Slater-TF} is valid for any values of $\mu$ as long as $\mu_k\ll1$. 
Also note that ${\rm arg}[a]\to\pi/2-\epsilon$ and ${\rm arg}[z]=\pi/2$. 
Thus, we have $2i a^{\frac{1}{2}} z^{\frac{1}{2}}=-x\left(1-3i\mu_k\right)^{1/2}$, hence the argument of trigonometric functions in \eqref{Slater-TF} becomes real and equal to $-x=-c_sk$ in the limit $\mu_k\ll1$.

On the other hand, in the limit $\mu\ll1$, since $x=\sqrt{\mu/\mu_k}$, the mode is always on superhorizon scales as long as  $\mu_k\gtrsim1$. Taking the limit $\mu\ll1$ of \eqref{t-transfer}, we find 
\begin{align}\label{TF-mu-small}
T_m(k,\tau)
\approx 1 - \frac{1}{6} x^2 - \frac{1}{20}\mu_k^2\,x^4\,; 
\qquad
x = \cs{k}\tau \,,
\quad
\mu=\frac{m}{H}\,.
\end{align}
Note that $\mu_k\,x^2=\mu\,(\ll 1)$.

Using the asymptotic forms \eqref{Slater-TF} and \eqref{TF-mu-small}, we can cover almost all possible regions of $\mu$ and $\mu_k$ which satisfy the relativistic condition \eqref{modes-R-mu}. 
We study the two cases, $\mu\ll1$ and $\mu\gg1$, separately below. 

\subsubsection{$m\ll{H}$ ($\Leftrightarrow \mu \ll 1$)}
As we have shown in appendix \ref{app-Kummer}, when $a$ takes the form \eqref{a-z-def} 
with $\mu_k \ll 1$ and $\mu\ll1$, \eqref{Slater-TF} simplifies to \eqref{Slater}.
Thus, the transfer function is given by
\begin{align}\label{t-transfer-small-mass}
T_m(k,\tau) \approx
\frac{\sin (x)}{x}
- \frac{\mu_k^2}{4}
\bigg[
\Big(x^2-1\Big) \frac{\sin (x)}{x}
- \left(\frac{2}{3}x^2-1\right)
\cos (x)
\bigg].
\end{align}
Note that $\mu_k^2\,x^2=\mu^2\ll1$. Hence the terms proportional to $\mu_k^2$ is always a small correction.
In the massless limit $m=0$, we have $\mu_k=0$ and recover the well-known result,
\begin{align}\label{t-transfer-massless}
T_0(k,\tau) = \frac{\sin (x)}{x} \,.
\end{align}
This is often used to study the time evolution of the primordial gravitational waves \cite{Watanabe:2006qe}.

When $\mu_k >1$, which corresponds to the superhorizon regime since $x^2=\mu/\mu_k\ll1$, the transfer function reduces to \eqref{TF-mu-small},
\begin{align}\label{TF-Rel-superhorizon}
T_m(k,\tau)
\approx 1 - \frac{1}{6} x^2 \Big( 1 + \frac{3}{10}\mu_k^2\,x^2 \Big) \,, 
\qquad
x\ll1 \,.
\end{align}


\subsubsection{$m\gg{H}$ ($\Leftrightarrow \mu \gg 1$)}
For the large mass limit $\mu\gg1$, we cannot terminate the series in \eqref{Slater-TF} as terms with higher power in $\mu$ give larger contributions. While there is no closed form for the series in general, in appendix \ref{app-Kummer} we have shown in detail that the series are convergent for $1\ll\mu\ll\mu_k^{-1}$. The result is given by Eq. \eqref{F-R-hz}. We note that, in this limit, $x\gg\mu\gg1$ such that
$\cs{k}/a \gg \m \gg H$. Thus, we can only use \eqref{F-R-hz} for the modes that are deep inside the horizon with mass much larger than the Hubble parameter. 
Picking up the leading terms in \eqref{F-R-hz}, we end up with 
\begin{align}\label{t-transfer-large-mass}
T_m(k,\tau) \approx
\frac{1 }{x} \sin \left( x+\frac{\mu_k^2}{6}x^3\right) \,, \qquad x \gg 1 \,.
\end{align}
Note that we cannot expand the argument of the $\sin$ function since $\mu_k^2\,x^3=\mu^2/x$ is not small. More precisely, the relativistic condition \eqref{modes-R-mu} only implies $\mu/x\ll1$, hence $\mu^2/x$ can be larger than unity.

\subsection{Non-relativistic regime}
The non-relativistic modes are those that have mass much larger than the physical momentum,
\begin{align}\label{NR-condition}
&\m \gg \frac{\cs{k}}{a} \,,
\qquad
\Rightarrow 
\qquad
\mu=\frac{m}{H} \gg \frac{H_k}{m}=\mu_k^{-1} \,.
\end{align}
Thus the non-relativistic condition \eqref{NR-condition} is rewritten as
\begin{align}\label{modes-NR-mu}
\mu\,\mu_k \gg 1 \,.
\end{align}For $\mu_k\gg1$ and $\mu\gg1$, this condition is trivially satisfied. 
However, when $\mu_k<1$ or $\mu<1$, the non-relativistic condition \eqref{modes-NR-mu} can only be satisfied if $\mu$ is sufficiently large (for $\mu_k < 1$) or $\mu_k$ is sufficiently
large (for $\mu <1$), respectively.

For $\mu\gg1$, there is an asymptotic series expansion for \eqref{t-transfer} as given by  \eqref{Kummer-Lz-p} in Appendix \ref{app-Kummer}. Setting $b=3/2$, which corresponds to our case, \eqref{Kummer-Lz-p} simplifies to
\begin{align}\label{t-transfer-Lz}
T_m(k,\tau)
&=
\frac{\sqrt{\pi}}{e^{\frac{\pi}{8\mu_k}} \mu^{\frac{3}{4}}}
{\rm Re}\left[
\frac{e^{\frac{i\mu}{2}-\frac{3i\pi}{8}} \mu^{\frac{i}{4\mu_k}}}{\Gamma\big(\frac{3}{4}+\frac{i}{4\mu_k}\big)}
\sum_{s=0}^{\infty}
\frac{\Gamma\big(\frac{1}{2}-\frac{i}{2\mu_k}+2s\big)}{\Gamma\big(\frac{1}{2}-\frac{i}{2\mu_k}\big)} \frac{\left(4i\mu\right)^{-s}}{s!}
\right] 
\,.
\end{align}

For $\mu_k \gg 1$, taking the limit $\mu_k\to\infty$ of \eqref{t-transfer}, 
we find
\begin{align}\label{t-transfer-NR-small-muk}
T_m(k,\tau) \approx \sqrt{2} \Gamma \Big(\frac{5}{4}\Big)
\mu^{-1/4} J_{\frac{1}{4}}\left(\frac{\mu}{2}\right) \,,
\end{align}
where $J_n(x)$ is the Bessel function of the first kind.

\subsubsection{$m\ll{H}$ ($\Leftrightarrow \mu \ll 1$)}
In the limit $\mu\ll1$, the non-relativistic condition \eqref{modes-NR-mu} implies $\mu_k\gg1$, hence the modes are on superhorizon scales. 
Taking $\mu\ll1$ limit of \eqref{t-transfer-NR-small-muk}, we find
\begin{align}\label{t-transfer-NR-S}
T_m(k,\tau) \approx 1 - \frac{\mu^2}{20}=1-\frac{\mu_k^2}{20}x^4 \,;
\quad 
x\ll 1 \,.
\end{align}

\subsubsection{$m\gg{H}$ ($\Leftrightarrow \mu \gg 1$)}
Now, we look at the regime $\mu\gg1$ when the mass dominates over not only the physical momentum but also the Hubble parameter. Using the asymptotic formula \eqref{t-transfer-Lz}, we find
\begin{align}\label{t-transfer-NR}
T_m(k,\tau) \approx \sqrt{\pi} e^{-\frac{\pi}{8\mu_k}} \mu^{-3/4}
{\rm Re}\left[
\frac{e^{\frac{i\mu}{2}-\frac{3i\pi}{8}} \mu^{\frac{i}{4\mu_k}}}{\Gamma\big(\frac{3}{4}+\frac{i}{4\mu_k}\big)}
\left(1 -\frac{1}{4 \mu  \mu_k}+\frac{i}{16 \mu  \mu_k^2}-\frac{3i}{16 \mu }\right)
\right]
 \,.
\end{align} 
The $\mu^{-1}$ correction is trivially suppressed since $\mu\gg1$. The $\mu^{-1}\mu_k^{-1}$ correction is also suppressed due to the  non-relativistic condition \eqref{modes-NR-mu}. However, the term $\mu^{-1}\mu_k^{-2}$ may not be necessarily suppressed since the condition $\mu\mu_k^2\gg1$ is stronger than the non-relativistic condition \eqref{modes-NR-mu} if $\mu_k<1$. 
Nevertheless, it is found to give a correction only to the phase when $\mu_k<1$, hence does not change our final result.\footnote{To see this explicitly, we note that $\Gamma\big(\frac{1}{2}-\frac{i}{2\mu_k}+2s\big)/\Gamma\big(\frac{1}{2}-\frac{i}{2\mu_k}\big)\approx\left(-4\mu_k^2\right)^{-s}$ for $\mu_k\ll1$, which after substituting in the summation in Eq.~\eqref{t-transfer-Lz} gives $\sum_{s=0}^{\infty}
\left(-4\mu_k^2\right)^{-s} \frac{\left(4i\mu\right)^{-s}}{s!}=\exp\left(\frac{i}{16\mu\mu_k^2}\right)$. Thus, the term $\frac{i}{16\mu\mu_k^2}$ in Eq.~\eqref{t-transfer-NR} is suppressed for $\mu_k>1$ and gives a correction only to the phase when $\mu_k<1$.}
The result of this subsection can be used for both subhorizon $x>1$ and superhorizon $x<1$ modes.


\section{Energy density}
\label{sec:energy-density}

The spectral energy density of $X$ is given by Eq. \eqref{Omega-def}. Substituting \eqref{transfer-function-def} we find
\begin{align}\label{Omega-def-final}
\Omega_X(k,\tau) 
&=
\frac{1}{a^2H^2}
\left[ 
| T'_m|^2 + \left( \cs^2 k^2 + \m^2 a^2 \right) | T_m |^2
\right] 
\frac{{\cal P}_{X,i}(k)}{6\Mpl^2}
\,,
\end{align}
where ${\cal P}_{X,i}(k)$ is the initial power spectrum that is defined in \eqref{PS-inf} and time $\tau$ denotes any time during radiation dominance. 
Note that we have set $f=1$ after inflation. Substituting \eqref{t-transfer-0} in \eqref{Omega-def-final} we can find a general explicit expression for the spectral energy density. However, such an expression in terms of Kummer's function is not very useful for practical purposes. In the next subsections, using the limiting forms of the transfer function for the relativistic and non-relativistic modes found in the previous section, we derive simple expressions for the spectral energy density.

\subsection{Relativistic regime}
The relativistic modes satisfy $m^2/(HH_k)=\mu\mu_k\ll1$ as given in
\eqref{modes-R} or \eqref{modes-R-mu}. 
We consider two regimes of $m\ll{H}$ and $m\gg{H}$ separately.

\subsubsection{$m\ll{H}$ ($\Leftrightarrow\mu\ll1$)}
In regime $\mu\ll1$ and $\mu\mu_k\ll1$, the transfer function is given by \eqref{t-transfer-small-mass}.
After substituting it in \eqref{Omega-def-final} we obtain
\begin{align}\label{Omega-R-S-split}
\begin{split}
\Omega_X(k,\tau)
&= \frac{{\cal P}_{X,i}(k)}{6\Mpl^2}
\begin{cases}
x^2 \left( 1 + \mu_k^2\,x^2 \right)\,;
&x < 1\,,
\\
1 - \dfrac{\sin(2x)}{x} \left(1 + \dfrac{1}{3}\mu_k^2\,x^2 \right) 
+ \dfrac{1}{6} \mu_k^2\,x^2 \cos(2x)\,;
&x \gg 1\,,
\end{cases}
\end{split}
\end{align}
where $x=\sqrt{\mu/\mu_k} = \cs k \tau$ as before, hence $\mu_k^2x^2\ll1$.

Thus, up to some small mass corrections, the homogeneous dimensionless energy density \eqref{Omega-homogeneous} for the relativistic modes in radiation dominated era is
\begin{align}\label{Omega-R-S-split-BG}
\begin{split}
{\bar \Omega}_X(\tau)
&\propto
\begin{cases}
a^2\,;
&x < 1\,,
\\
a^0\,;
&x \gg 1 \,.
\end{cases}
\end{split}
\end{align}
For $x < 1$, the gradient term dominates, and $\bar{\Omega}_X(\tau)$ behaves like spatial curvature. In contrast, for $x \gg 1$, both kinetic and gradient terms contribute equally, and $\bar{\Omega}_X(\tau)$ behaves like radiation, as expected. In the former case, there may be interesting features, which are beyond the scope of this paper. In the latter case, the additional dark radiation would contribute to the effective number of relativistic degrees of freedom~\cite{Planck:2018vyg}, leading to constraints from big bang nucleosynthesis (BBN).

\subsubsection{$m\gg{H}$ ($\Leftrightarrow\mu\gg 1$)}
As we have already mentioned, $\mu\gg1$ together with the relativistic condition \eqref{modes-R-mu} implies $x\gg1$ which means we only deal with the modes deep inside the horizon. Substituting the transfer function \eqref{t-transfer-large-mass} in \eqref{Omega-def-final} gives
\begin{align}\label{Omega-R-L}
\Omega_X(k,\tau) \approx
\frac{{\cal P}_{X,i}(k)}{6\Mpl^2} \,,
\qquad x\gg1 \,.
\end{align}
Thus, in this case, we find
\begin{align}\label{Omega-R-S-split-BG-m}
\begin{split}
\bar{\Omega}_X(\tau)
\propto
a^0\,;
\qquad
x \gg 1 \,,
\end{split}
\end{align}
which shows that the averaged energy density behaves like radiation. This is because, for modes deep inside the horizon ($x \gg 1$), both kinetic and gradient terms contribute equally. Similar to the $x \gg 1$ case in \eqref{Omega-R-S-split-BG}, there will be a BBN constraint on the energy density.

\subsection{Non-relativistic regime}
The non-relativistic regime satisfies  $m^2/(HH_k)=\mu\mu_k\ll1$ as
given by \eqref{NR-condition} or  \eqref{modes-NR-mu}. 
Similar to the previous subsection, we consider two regimes of $m\ll{H}$ and $m\gg{H}$ separately.

\subsubsection{$m\ll{H}$ ($\Leftrightarrow\mu\ll1$)}
The non-relativistic limit $\mu\mu_k\gg1$ in \eqref{modes-NR-mu} together with $\mu\ll1$ implies $x=c_sk\tau\ll1$ which means that we only deal with the superhorizon modes. 
The transfer function is given by \eqref{t-transfer-NR-S} which after substituting in  \eqref{Omega-def-final} gives
\begin{align}\label{Omega-NR-S}
\Omega_X(k,\tau)
=
\mu^2 \Big(1 + \frac{1}{\mu\mu_k} \Big) \frac{{\cal P}_{X,i}(k)}{6\Mpl^2} \,;
\quad
x \ll 1
\,,
\end{align}
where $\mu = m/H$ and $\mu_k={m}/{H_k}$ as before.
For the time dependency of the homogeneous energy density \eqref{Omega-homogeneous}, we find 
\begin{align}\label{Omega-NR-S-BG}
{\bar \Omega}_X(\tau)
\propto
a^4 \,;\quad
x \ll 1
\,,
\end{align}
up to a small correction proportional to the inverse of mass squared. This means the energy density ${\bar \rho}_X(\tau)$ of these superhorizon modes is constant in time. This is because the mass term dominates, and therefore the energy density behaves like a cosmological constant.

\subsubsection{$m\gg{H}$ ($\Leftrightarrow \mu\gg1$)}
For the non-relativistic limit \eqref{modes-NR-mu} with $\mu\gg1$, the mass term dominates both the physical momentum and the Hubble parameter. The transfer function is given by \eqref{t-transfer-NR}. Substituting it in \eqref{Omega-def-final} gives
\begin{align}\label{Omega-eq-limit}
\Omega_X(k,\tau) \approx
\frac{4\pi \mu^{1/2} e^{-\frac{\pi}{4\mu_k}}}{
\big{|}\Gamma \big(\frac{3}{4} + \frac{i}{4\mu_k}\big)\big{|}^2}
\frac{{\cal P}_{X,i}(k)}{6\Mpl^2}
\,,
\end{align}
where we have neglected subleading mass corrections.

For the homogeneous dimensionless energy density we find
\begin{align}\label{Omega-eq-limit-BG}
{\bar \Omega}_X(\tau) \propto a \,,
\qquad
{\bar \rho}_X(\tau) \propto a^{-3} \,,
\end{align}
which shows that the energy behaves like dust. This is because the kinetic and mass terms dominate equally. Note that this result holds for both subhorizon and superhorizon modes. This is indeed the case of interest in this paper, as $X$ can play the role of dark matter, which is the subject of the next section.

\section{Dark matter}
\label{sec:dark-matter}
Up to this point, we have derived the analytical transfer function and energy density for the massive field $X$, with initial conditions set by the modes stretched to superhorizon scales during inflation which subsequently re-enter the horizon in the radiation dominated era. 
During the whole process (both inflationary and radiation eras), the field $X$ was assumed to be a spectator field. 
In this section, we assume that $X$ dominates at the time of the matter-radiation equality and plays the role of dark matter.

\subsection{Amplitude of the dark matter density contrast}
We deal with the quantum fluctuations of $X$ where the homogeneous, zero mode component is negligible, as given by the condition \eqref{X-vev}. 
Then, the mean energy density of $X$ is given by the vacuum average \eqref{rho}, which describes the accumulative energy of the excited quantum fluctuations. 
By the time of the matter-radiation equality, we require the mass term in \eqref{rho} dominates over
the gradient term (non-relativistic regime) and $m\gg H$ so that $X$ plays the role of dark matter.
In this regime, we find 
\begin{align}
{\bar \rho}_X &\approx \frac{1}{2} {m}^2\sigma_X^2 \,,
\end{align}
where $\sigma_X^2=\langle {\hat X}^2 \rangle - \langle {\hat X} \rangle^2
\approx \langle {\hat X}^2 \rangle$ as defined in \eqref{X-vev}. 
Using the Gaussianity of $\hat X$, the variance of the dark matter density contrast
\begin{align}\label{delta}
\delta_X
\equiv\frac{\rho_X-{\bar \rho}_X}{\bar \rho_X} \,,
\end{align}
is computed as
\begin{equation}\label{delta-V}
\langle \delta_X^2 \rangle =\bigg\langle {\left( \frac{\hat X^2-\sigma_X^2}{\sigma_X^2} \right)}^2 \bigg\rangle =2 \,.
\end{equation}
Detailed derivation of this result is given in appendix \ref{app-PS-DM}.
Thus, amplitudes of dark matter density contrast are typically ${\cal O}(1)$,
which is consistent with the result obtained in \cite{Alonso-Alvarez:2018tus}. 
Note also that even if $\hat X$ is Gaussian, e.g. $\langle{\hat X}^3\rangle=0$, the distribution of the dark matter density contrast is highly non-Gaussian. For instance, the skewness is given by
\begin{equation}\label{NG}
\langle \delta_X^3 \rangle = 8 \,.
\end{equation}
Clearly, if the scale of fluctuations is at the scales probed by CMB and LSS observations,
such large inhomogeneities are inconsistent with the fact that the universe
is very close to the FLRW universe. 

From now on, we set the sound speed to be unity $\cs=1$ for the sake of simplicity. As the CMB and LSS scales are more or less around $k_{\rm eq}$, where $k_{\rm eq}\sim{10}^{-2}\mbox{Mpc}^{-1}$ is the comoving wavenumber which re-enters the horizon at the time of matter-radiation equality, 
we should restrict our scenario to the case where the scale-dependent initial power spectrum with $f\neq1$ excludes excitations of the modes $X_k$ with $k\lesssim{k}_{\rm eq}$.
Actually, in this case, the dimensionless power spectrum of $\delta_X$ is extremely blue on large scales. It scales as ${\cal P}_{\delta_X} \propto k^3$ in general, except for the case of a monochromatic power spectrum proportional to Dirac's delta function in which $\propto k^2$.
Thus, adiabatic perturbations coming
from the inflaton fluctuations, which are nearly scale-invariant, become a dominant component of dark
matter perturbations on large scales and there is no deviation from the $\Lambda$CDM model predictions on those scales.\footnote{
Adiabatic perturbations from the inflaton are imprinted in the perturbations of $X$ at the onset of its oscillations. This is because the starting time of the oscillations is determined by the
uniform density hypersurface ($H=m$) on which the perturbation of radiation density vanishes.}

The freedom in choosing $f$ makes it possible to have initial power spectra with different shapes. In this paper, as the simplest case, we choose an initial power spectrum which has a sharp peak at small scales $k\gg{k}_{\rm eq}$. This toy model not only significantly simplifies the setup but also clarifies some universal features of our dark matter scenario. We consider the case that $f$ is chosen such that the initial power spectrum has a sharp peak at scale $k_p=-1/\tau_p$ while the power spectrum is completely suppressed at any other scales. 
Hence, we approximate the initial power spectrum to have the monochromatic form,
\begin{align}\label{PS-delta}
{\cal P}_{X,i}(k) = {\cal A} \left(\frac{H_{\rm inf}}{2\pi}\right)^2 \delta\left[\ln\left(k/k_p\right)\right] \,;
\qquad 
k_p \gg k_{\rm eq} \,,
\end{align}
where ${\cal A}$ is a free dimensionless coefficient. Note that ${\cal A}$ can be very large such that the power spectrum is only enhanced at $k=k_p$ and completely suppressed at other scales \cite{Pi:2021dft}. Thus, in this simplest setup, there are three free parameters in the model;
the mass $m$, the scale $k_p$, and the amplitude ${\cal A}$.


\subsection{Relic density}
Dark matter particles are non-relativistic such that the mass dominates both the physical momenta and Hubble parameter. In this case, the energy density is given by \eqref{Omega-eq-limit}. 
Substituting \eqref{PS-delta} in \eqref{Omega-eq-limit} gives
\begin{align}\label{Omega-delta}
&\Omega_X(k,\tau) = \frac{1}{24} \beta_{\rm inf} \left(\frac{\m}{H}\right)^{1/2}
|{\cal I}(k)|^2 \delta\left[\ln\left(k/k_p\right)\right] 
\,,
\end{align}
where we have defined time-independent function
\begin{align}\label{I-def}
{\cal I}(k) \equiv \sqrt{\frac{\pi}{2}}
\frac{e^{-\frac{\pi H_k}{8\m}}}{
	\Gamma \left(\frac{3}{4} + \frac{iH_k}{4\m}\right)} \,,
\end{align}
which only depends on the scale, and 
\begin{align}\label{beta-c-defs}
&\beta_{\rm inf} \equiv 
r\, {\cal A} \,{\cal P}_{\cal R}\,, 
\end{align}
in which $r={\cal P}_h/{\cal P}_{\cal R}$ is the tensor-to-scalar ratio, ${\cal P}_h=2H_{\rm inf}^2/(\pi^2\Mpl^2)$
and ${\cal P}_{\cal R}$ are the power spectra of the primordial gravitational waves and the curvature perturbation, respectively. 
Note that since $X$ is a spectator field during inflation, it does not significantly contribute to the metric tensor perturbation and curvature perturbation. 
Thus, ${\cal P}_h$ and ${\cal P}_{\cal R}$ are the same as the usual power spectra in single field inflation. 
However, in the case that $X$ is a spin-2 field, there can be a significant contribution to ${\cal P}_h$ \cite{Gorji:2023ziy,Gorji:2023sil}. 
In this case, one needs to carefully take into account the contribution of the field $X$ as a source of gravitational waves.

\begin{figure}[!ht]
	\centering
	\includegraphics[width=.8 \columnwidth]{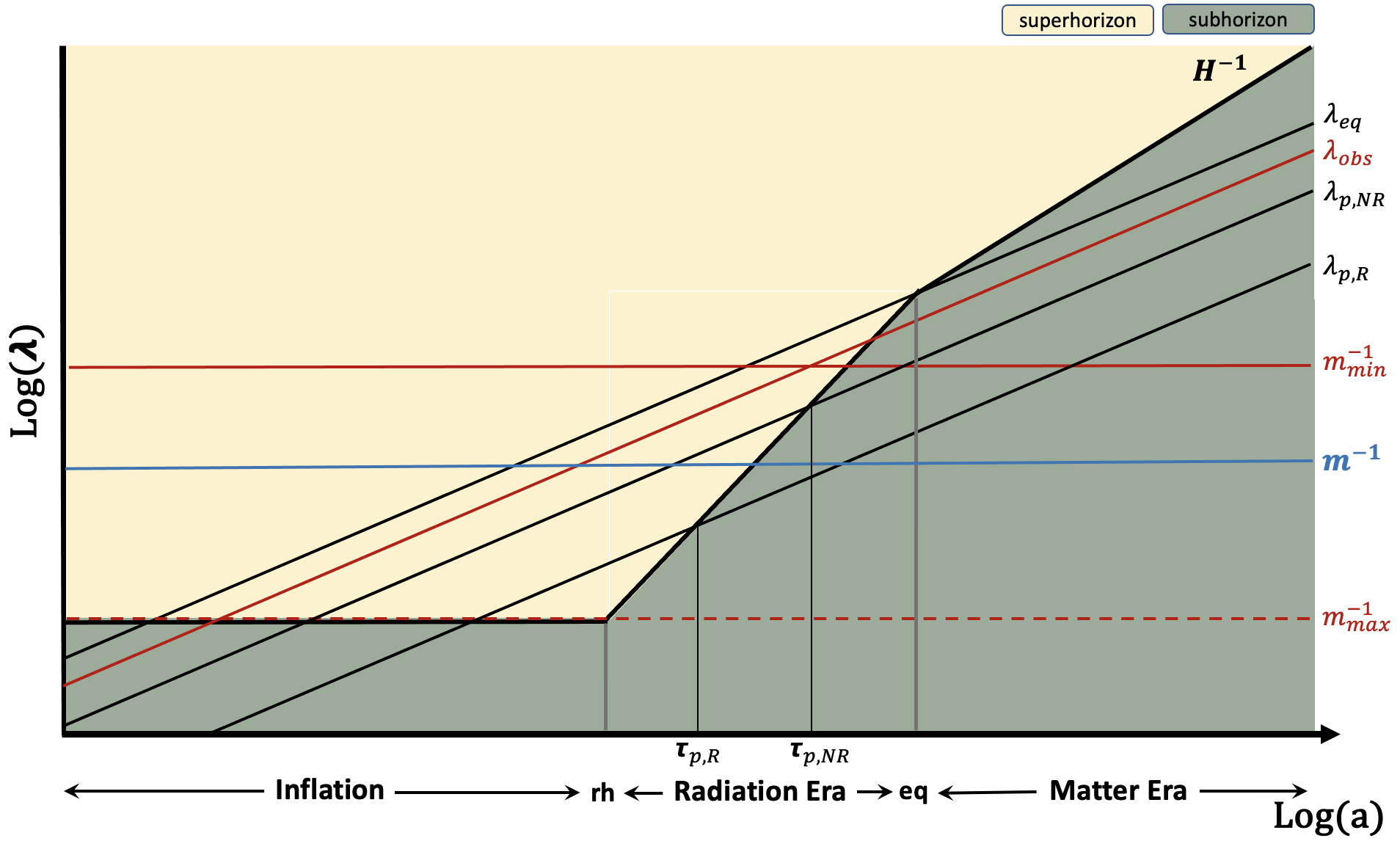}
	\caption{Physical scale $\lambda\sim a/{k}$ versus the scale factor in logarithmic scales are plotted. The lower $m_{ min}^{-1}$ and upper $m_{ max}^{-1}$ bounds on the mass are defined by Eqs. \eqref{mass-lower-bound} and \eqref{mass-upper-bound} respectively. Avoiding large fluctuations at $\lambda\gtrsim\lambda_{\rm obs}$ to be consistent with CMB and LSS observations, we have to restrict our setup to the scales $\lambda\lesssim\lambda_{\rm obs}$. The blue solid line $m^{-1}$ represents a typical mass value which should lie in the range $m_{min}\lesssim{m}\lesssim{m}_{max}$. The solid lines $\lambda_{p,{\rm R}}\sim a/{k}_{p,{\rm R}}$ and $\lambda_{p,{\rm NR}}\sim a/{k}_{p,{\rm NR}}$ show two possible momenta $k_{p,{\rm R}}$ and $k_{p,{\rm NR}}$ at which the initial power spectrum can have a peak. The mode $k_{p,{\rm R}}$ is relativistic at the time of horizon re-entry since $\lambda_{p,{\rm R}}(\tau_{p,{\rm R}})<m^{-1}$ or $k_{p,{\rm R}}/a(\tau_{p,{\rm R}})>m$ while $k_{p,{\rm NR}}$ is non-relativistic since $\lambda_{p,{\rm NR}}(\tau_{p,{\rm R}})>m^{-1}$ or $k_{p,{\rm NR}}/a(\tau_{p,{\rm R}})<m$. The latter case, which corresponds to $m>H_{p,{\rm NR}}$, collapses to form subsolar mass dark matter halos at high redshifts.
	}
	\label{fig-modes}
\end{figure}

\begin{figure}[!ht]
	\centering
	\includegraphics[width=.7 \columnwidth]{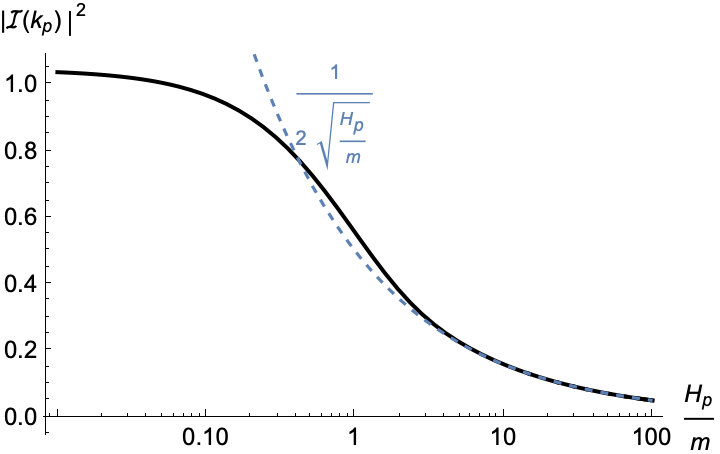}
	\caption{The function $|{\cal I}(k_p)|^2$ is monotonically decreasing as mass decreases and it is bounded as $0<|{\cal I}(k_p)|^2\leq \pi/[2\Gamma(3/4)^2]$. The tail for light masses $m\ll{H}_p$ is well described by $1/(2\sqrt{H_p/m})$. Moreover, up to masses $m\sim{H}_p$ can be approximated with the tail behavior $1/(2\sqrt{H_p/m})$ with error less than ${\cal O}\left(10^{-3}\right)$. For heavy masses $m\gg{H}_p$, the function is almost independent of the mass and approaches to $\pi/[2\Gamma(3/4)^2]\approx1.046$.}
	\label{fig-I}
\end{figure}

For later convenience, we provide an approximate expression of ${\cal I}(k)$ in the two limiting cases: i) $H_p\gtrsim{m}$, and ii) $H_p\ll{m}$
where we have defined
\begin{align}\label{Hp}
&H_p \equiv H_{k_p} = \frac{{k}_p}{a_p} \,.
\end{align}
The case i), represented by $\lambda_{p,{\rm R}}$ in Fig.\ref{fig-modes}, corresponds to the modes which are relativistic at the time of horizon re-entry, while the case ii), represented by $\lambda_{p,{\rm NR}}$ in Fig.\ref{fig-modes}, refers to the modes which are non-relativistic at the time of horizon re-entry. These two cases cover almost entire parameter space of interest, and the function ${\cal I}(k)$ defined in \eqref{I-def} takes very simple forms,
\begin{align}\label{I-expansion}
{\cal I}(k_p) = 
\begin{cases}
\dfrac{1}{\sqrt{2}} \big(\dfrac{\m}{H_p}\big)^{1/4}
\,; 
&\m\lesssim{H}_p \,,
\\
\sqrt{\dfrac{\pi }{2}}\dfrac{1}{\Gamma \left(3/4\right)}
\,;
&\m\gg{H}_p \,.
\end{cases}
\end{align}
Here we have used the identity $\lim_{|a|\to\infty} |\Gamma(b+ia)|= \sqrt{2\pi} |a|^{b-1/2}e^{-\pi|a|/2}$. The function $|{\cal I}(k_p)|^2$ is depicted in Fig.~\ref{fig-I}.
As it can be seen, $|{\cal I}(k_p)|^2$ is a monotonic function of mass. It is almost independent of the mass for $m\gg{H}_p$ where its numerical value is almost unity $\pi/[2\Gamma(3/4)^2]\approx1.046$, while it monotonically decreases when the mass becomes smaller and smaller and it asymptotically approaches zero for $m\to0$. 
The tail $m\ll{H}_p$ can be very well approximated by the function $1/(2\sqrt{H_p/m})$.  
Moreover, this approximation is accurate enough up to the mass $m\sim{H}_p$ with an error less than ${\cal O}\left(10^{-3}\right)$. Thus, we can safely use $1/(2\sqrt{H_p/m})$ for the whole range of $m\lesssim{H}_p$.

Using \eqref{I-expansion} for $m\lesssim{H}_p$ in \eqref{Omega-delta} and then substituting the result in \eqref{Omega-homogeneous} we find
\begin{align}\label{Omega-delta-Light}
&{\bar \Omega}_X(\tau) = \frac{1}{48} \beta_{\rm inf} \left(\frac{\m}{H}\right)^{1/2}
\left(\frac{\m}{H_p}\right)^{1/2}
\,,
\qquad m\lesssim{H}_p \,.
\end{align} 
For the modes $m \gg {H}_p$, using \eqref{I-expansion} for $m\gg{H}_p$ in \eqref{Omega-delta} and substituting the result in \eqref{Omega-homogeneous}, we find
\begin{align}\label{Omega-delta-Heavy}
&{\bar \Omega}_X(\tau) = \frac{\pi}{48\Gamma(3/4)^2} \beta_{\rm inf} \left(\frac{\m}{H}\right)^{1/2}
\,,
\qquad
m \gg {H}_p \,.
\end{align}

\subsection{Constraints on the model parameters}

Our model contains three free parameters $(m, k_p, \beta_{\rm inf})$. 
As we have already mentioned, there is an upper bound on the mass of the dark matter particles 
given by Eq. \eqref{massless-inf}. 
This relation can be rewritten in the following more appropriate form,
\begin{align}\label{mass-upper-bound}
m \ll H_{\rm inf} \sim  10^{13}\,\mbox{GeV}~\left(\frac{r}{0.03}\right)^{1/2} \left(\frac{{\cal P}_{\cal R}}{3\times{10}^{-10}}\right)^{1/2} \,. 
\end{align}
On the other hand, finding a lower bound on the mass needs careful considerations since it completely depends on the scale of our interest.
However, as we will show, under the assumption of a sharply peaked power spectrum at $k_p$, it can be done in a simple manner.
In what follows, in addition to the above constraint on the mass,
we give four conditions based on the cosmological observations that support 
the standard $\Lambda$CDM model. 
We will use them to identify the allowed region in the parameter space.

\begin{figure}[t]
	\centering
	\includegraphics[width=.8 \columnwidth]{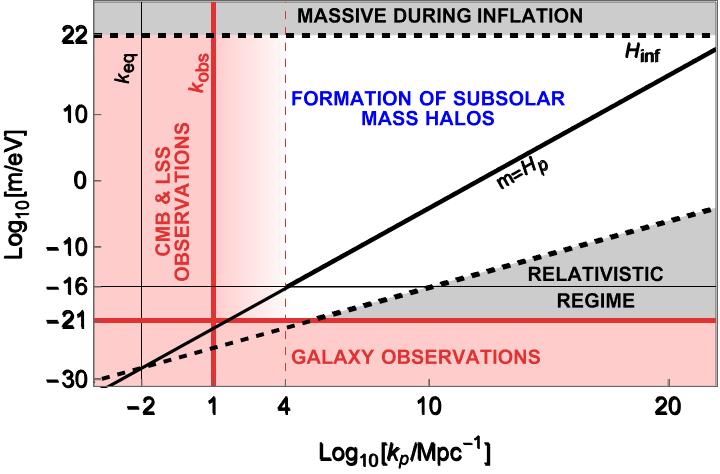}
	\caption{Mass of dark matter versus the scale is plotted. The red regions are excluded by the observations while the gray regions are excluded by the theoretical constraints. For the upper bound on the mass $m\ll{10}^{22}\,\mbox{eV}$, we have considered $H_{\rm inf}=10^{13}\,\mbox{GeV}$ in \eqref{mass-upper-bound} which corresponds to $r\sim0.03$. For lower values of tensor-to-scalar ratio, this bound becomes tighter. The lower bound on the mass $m \gtrsim 10^{-21}\,\mbox{eV}$ comes from the galaxy observations. The lower bound on the scale $k_p \gtrsim 10^3{k}_{\rm obs}\sim 10\,\mbox{kpc}^{-1}$ should be imposed to be consistent with the absence of large fluctuations in the CMB and LSS observations. The allowed region restricted to $m>H_p$, that corresponds to the modes which are non-relativistic at the time of horizon re-entry, will form subsolar mass dark matter halos at high redshifts.}
	\label{fig-constraints}
\end{figure}

\begin{itemize}
	\item{\bf $X$ as all dark matter:} Our assumption is that the $X$ field comprises
		all the dark matter. This means that at the time of matter-radiation equality, we should require
	\begin{align}\label{Omega-eq}
	{\bar \Omega}_X (\tau_{\rm eq})=\frac{1}{2}.
	\end{align}
	\item {\bf Non-relativistic condition:} The dark matter particles should be non-relativistic before the time of matter-radiation equality:
  ${k_p}/{a_{\rm eq}}\leq \epsilon\, m$ with $\epsilon \ll 1$. 
 Here $\epsilon$ represents how non-relativistic the dark matter should be at the time of matter-radiation equality.
 Although there are studies on how much dark matter can deviate from the CDM in the early universe, there is uncertainty in the upper limit on $\epsilon$ consistent with cosmological observations \cite{Kopp:2018zxp}.
 Here we simply adopt $\epsilon=0.1$ as a representative value. 
Thus, the mass should satisfy non-relativistic condition \eqref{NR-condition}. For the sharply peaked power spectrum at $k_p$ ($\cs=1$), it can be rewritten as
	\begin{align}\label{m-lower-bound}
	m \ge &
	\frac{1}{\epsilon} {\left(H_{\rm eq}H_p\right)}^{1/2}
	\approx
	2\times {10}^{-21}~{\rm eV} {\left( \frac{\epsilon}{0.1} \right)}^{-1}
	\left( \frac{k_p}{10~\mbox{kpc}^{-1}} \right)
	\,,
	\end{align}
	where we have substituted $H_{\rm eq}\approx 2\times 10^{-28}\,\mbox{eV}$ and $H_p$ is defined in Eq. \eqref{Hp}.
	
	\item {\bf CMB and LSS observations:} To be consistent with CMB and LSS observations, dark matter should follow the adiabatic fluctuations with scale-invariant power spectrum at large scales. The scale-invariant power spectrum has been observed up to the scales $k_{\rm obs}\lesssim\,10\mbox{Mpc}^{-1}$.
	Thus, we require that the isocurvature power spectrum is smaller than the adiabatic one on those scales. 
    Adopting the extremely blue spectrum ${\cal P}_{\delta_X} ={(k/k_p)}^3$, the condition
   ${\cal P}_{\delta_X} ={(k_{\rm obs}/k_p)}^3 \lesssim 10^{-10}$ yields
	\begin{align}\label{scale-lower-bound-CMB}
	k_p \gtrsim 10^3 {k}_{\rm obs}\sim 10\,\mbox{kpc}^{-1}\,.
	\end{align}
	The above bound is independent of the spin of particles and it is solely based on the argument of avoiding large fluctuations \eqref{delta-V} and \eqref{NG} at the scales $k_p<{k}_{\rm obs}$. 
	
	\item {\bf Galactic structures:} 
 There are various ways to constrain the mass of ultra-light dark matter from observations of galactic structures such as density profile, satellite mass, and satellite abundance. 
 Precise values of the lower limit on the mass depend on the methods, but they are more or less 
 $10^{-21}~{\rm eV}$ (see \cite{Hui:2021tkt} and references therein). Lyman-alpha forest, which is the spectrum of a distant source and is caused by clouds of neutral hydrogen between the Earth and the source, is another useful probe of dark matter mass and provides similar lower limit on the mass \cite{Irsic:2017yje, Kobayashi:2017jcf, Armengaud:2017nkf}.
 In this paper, we adopt
	\begin{align}\label{mass-lower-bound}
	m \gtrsim 10^{-21}\,\mbox{eV} \,,
	\end{align}
as a representative value of the lower limit. It is interesting to note that for the largest scale $k_p\sim 10\,\mbox{kpc}^{-1}$  allowed by the CMB and LSS observations given in \eqref{scale-lower-bound-CMB}, the non-relativistic condition \eqref{m-lower-bound} coincides with the above lower bound on mass (see also Fig.~\ref{fig-constraints}).
	
\end{itemize}

For any given mass $m$ and peak scale $k_p$ in the allowed region, one can always substitute \eqref{Omega-delta} in \eqref{Omega-eq} to find the corresponding $\beta_{\rm inf}$. 
This may be performed numerically, but for the two limiting cases $m\lesssim H_p$ and
$m \gg H_p$, $\beta_{\rm inf}$ can be easily obtained as follows.
For $m\lesssim H_p$, using Eq.~(\ref{Omega-delta-Light}) and the relation
$H_p=k_p^2/(a_{\rm eq}^2 H_{\rm eq})$, we obtain
\begin{equation}
{\bar \Omega_X}(\tau_{\rm eq})=\frac{1}{48}\beta_{\rm inf} \frac{m}{k_p}a_{\rm eq}\,.
\end{equation}
Thus, the condition ${\bar \Omega_X}(\tau_{\rm eq})=\frac{1}{2}$ fixes $\beta_{\rm inf}$ as
\begin{equation}
\label{beta-inf-1}
\beta_{\rm inf}=24 \frac{k_p}{a_{\rm eq} m}\,.
\end{equation}
We find that the non-relativistic condition \eqref{m-lower-bound} gives $\beta_{\rm inf}\ll 24$.
For $m \gg H_p$, using Eq.~(\ref{Omega-delta-Heavy}) and imposing the condition ${\bar \Omega_X}(\tau_{\rm eq})=\frac{1}{2}$, we obtain
\begin{equation}
\beta_{\rm inf}=\frac{24 \Gamma (\frac{3}{4})}{\pi} 
{\left( \frac{H_{\rm eq}}{m} \right)}^\frac{1}{2}.
\end{equation}

After fixing $\beta_{\rm inf}$ in this way, the remaining parameters are $m$ and $k_p$.
We have plotted the allowed region (white region) for $m$ and $k_p$ in Fig.~\ref{fig-constraints}, where we have imposed both observational \eqref{scale-lower-bound-CMB}, \eqref{mass-lower-bound} (red regions) and theoretical \eqref{mass-upper-bound}, and \eqref{m-lower-bound} constraints (gray regions).
The black thick line represents $m=H_p$.
The dashed line dividing non-relativistic and relativistic regime is defined as 
$\frac{k}{a_{\rm eq}}=\epsilon H_{\rm eq}$ with $\epsilon=0.1$.
We see that there is a vast range of the allowed region both for $m>H_p$ and $m <H_p$. 

\subsection{Halo formation}
Although any point in the allowed region is equally allowed at the level of the conditions we have imposed, our model may be tested against future observations, since it predicts characteristic inhomogeneities on small scales that do not exist in the standard $\Lambda$CDM model.
One notable feature is that the dark matter density contrast at the scale $k_p$ has 
its variance $\langle \delta_X^2 \rangle =2$ from the outset.
This magnitude is much larger than the one $\sim 10^{-10}$ observed at CMB and LSS scales.
Such large dark matter fluctuations will result in the copious formation of dark matter halos 
in the early times even around the time of the matter-radiation equality.

In order to qualitatively estimate in which region in the parameter space this can happen, 
let us compare the two time scales, the sound crossing time $t_s$ and the free-fall time $t_f$, of an overdense region with its comoving size $1/k_p$. 
Because the proper wavenumber is $k_p/a$, the propagation speed of the wave is $\frac{k_p}{am}$.
Thus, the sound crossing time, which is the time the waves take to propagate over the size
$a/k_p$, becomes $t_s=\frac{a^2m}{k_p^2}$.
The free-fall time is $t_f=1/\sqrt{G\rho}$ ($\sim\Mpl/\sqrt{\rho}$ in our unit). Gravitational collapse of the overdense region will occur when $t_f<t_s$ is satisfied \cite{Khlopov:1985fch,Nambu:1989kh}.
Thus, the modes satisfying
\begin{equation}\label{kp-halo}
k_p < a(z) \sqrt{H (z) m} \simeq 1~{\rm kpc}^{-1} {\left( \frac{1+z}{100} \right)}^{-\frac{1}{4}}
{\left( \frac{m}{10^{-18}~{\rm eV}} \right)}^\frac{1}{2} \,,
\end{equation}
will produce dark matter halos by the cosmological redshift $z$. Using $k_p=k_{\rm eq}(H_p/H_{\rm eq})^{1/2}$, the above condition can be rewritten as
\begin{equation}\label{m-halo}
m > H_p {\left( \frac{1+z}{100} \right)}^{\frac{1}{2}} \,.
\end{equation}
After the time of matter-radiation equality, the redshift range is $0\leq{z}\lesssim3000$ which gives $0.1\leq{\left( \frac{1+z}{100} \right)}^{\frac{1}{2}}<5.5$. Thus, the condition to form dark matter halo is approximately defined by $m>H_p$ which corresponds to the modes which are non-relativistic at the time of horizon re-entry.

The typical mass of the dark matter halos is estimated as
\begin{equation}
M_h \simeq \frac{4\pi}{3} \rho_m k_p^{-3} \simeq 1.6\times 10^{2}~M_\odot~
{\left( \frac{k_p}{1~{\rm kpc}^{-1}} \right)}^{-3}.
\end{equation}
The CMB and LSS bound \eqref{scale-lower-bound-CMB} implies that
\begin{equation}
M_h < 1.6\times 10^{-1}~M_\odot \,.
\end{equation}
Thus, only subsolar-mass dark matter halos can form from the large isocurvature perturbations of \(X\). 
Applying the CMB and LSS bound \eqref{scale-lower-bound-CMB} to \eqref{kp-halo}, and using the lower bound on the mass \eqref{mass-lower-bound}, we find that within the mass range \(10^{-21}\,\mathrm{eV} \lesssim m < 10^{-16}\,\mathrm{eV}\), dark matter halos cannot form. For the mass range
\begin{align}\label{m-halos-lower}
	m \gtrsim 10^{-16}\,\mathrm{eV} \,, \qquad 
	\mbox{subsolar-mass halos form} \,,
\end{align}
subsolar-mass dark matter halos will form at high redshifts (see Fig.~\ref{fig-constraints}). Observational confirmation of a large population of such low-mass dark
matter halos would support our scenario in which dark matter
originates purely from quantum fluctuations during inflation. Here we
mention that the de Broglie length constraint on the halo formation
could further limit the mass range of our interest \cite{Ferreira:2020fam}.
However, a straightforward estimate tells us that the constraint \eqref{m-halos-lower} we
obtained, is already strong enough to satisfy the de Broglie length constraint.

\subsection{Big bang nucleosynthesis bound}
Before closing this section, we finally comment on the role of the $X$ field 
as dark radiation in the BBN era.
If the modes are subhorizon and relativistic, their energy density behaves like radiation. 
In the BBN era, this can happen for the parameter space simultaneously satisfying the conditions,
\begin{equation}
\frac{k_p}{a_{\rm BBN}} \gg m,
\qquad
\frac{k_p}{a_{\rm BBN}} \gg H_{\rm BBN}.
\end{equation}

The second condition is equivalent to 
\begin{equation}
k_p \gg 30~{\rm kpc}^{-1} \left( \frac{T_{\rm BBN}}{1~{\rm MeV}} \right),
\end{equation}
where $T_{\rm BBN}$ is the temperature of radiation at BBN era.
Since the radiation temperature varies by an order of magnitude ($0.1~{\rm MeV}\sim 1~{\rm MeV}$) throughout the BBN era, there is an ambiguity of choosing $T_{\rm BBN}$.
Here we take the largest temperature because this choice ensures that the modes remain subhorizon at any time during the BBN era. 
Imposing this condition, we find that the parameter space where the
first condition is met is in the regime $m\lesssim H_p$.
Then, from Eq.~(\ref{beta-inf-1}), $\beta_{\rm inf}$ on the boundary between
the non-relativistic and the relativistic regime in Fig.~\ref{fig-constraints}
(i.e., $\frac{k_p}{a_{\rm eq}}=\epsilon m$)
becomes
\begin{equation}
\beta_{\rm inf}=24\epsilon\,.
\end{equation}
On the other hand, when the modes are relativistic and in the regime $m\lesssim H_p$, we have 
$\Omega_X(k,\tau) \simeq {\cal P}_{X,i}(k)/(6\Mpl^2)$ from 
\eqref{Omega-R-L}. 
Using the assumed power spectrum given by \eqref{PS-delta}, we find
\begin{align}\label{Omega-delta-Light-R}
&{\bar \Omega}_{X} (\tau_{\rm BBN}) = \frac{1}{48} \beta_{\rm inf}\,.
\end{align} 

The contribution to the effective number of relativistic degrees of freedom is given by $\Delta{N}_{\rm eff} = (8/7) \left(11/4\right)^{4/3} {\bar \Omega}_X$. Applying the observational bound by Planck $\Delta{N}_{\rm eff}<0.3$ \cite{Planck:2018vyg} to it implies
\begin{align}\label{BBN-bound} 
\beta_{\rm inf} < 3.27.
\end{align}
Thus, the BBN bound \eqref{BBN-bound} is satisfied for $\epsilon=0.1$. Note, however, the BBN bound may become relevant for a marginally non-relativistic case if larger value of $\epsilon$ is adopted.




\section{Summary and discussion}\label{summary}

We studied a scenario in which dark matter is a massive bosonic field, arising solely from the quantum fluctuations that are stretched to the superhorizon scales during inflation. In this scenario, dark matter exhibits ${\cal O}(1)$ primordial isocurvature perturbations at small scales that are beyond the reach of current observations, such as those from the CMB and LSS. Neglecting changes in the number of relativistic degrees of freedom, we derived an exact transfer function for the dark matter field perturbations during the radiation dominated era. 
Depending on the hierarchy between the mass $m$, the scale $k$, and the Hubble parameter $H$, the perturbations show distinct behaviors across different regimes. We derived approximate expressions for the transfer function for each regime and reproduced the known behaviors discussed in the literature. 
Assuming a monochromatic initial power spectrum with a peak at $k=k_p$, we identified the viable parameter space in the $m$-$k$ plane as $10^{-21}\,\mbox{eV}\lesssim{\m}\ll H_{\rm inf}$ and $k_p\gtrsim10~{\rm kpc}^{-1}$. The lower bound on the mass comes from the observation of the galactic structures while the upper bound should be imposed to have effectively massless perturbations during inflation such that they freeze at the superhorizon scales. 
The lower bound on the scale $k_p$ is to avoid large fluctuations, e.g. large isocurvature perturbations, on large scales probed already by the CMB and LSS observations.

A key prediction of this scenario is the abundant formation of subsolar mass dark matter halos at high redshifts due to ${\cal O}(1)$ fluctuations at small scales. Our result shows that this happens for the modes that are non-relativistic at the time of horizon re-entry during radiation domination. Observational confirmation of a large population of such low-mass halos will support the hypothesis that dark matter is originated purely from inflationary quantum fluctuations of a massive bosonic field.

Finally, let us mention possible mechanisms to realize our scenario. The most important assumption of our scenario is that the field $X$ is light during inflation, i.e.,
$m\ll H_{\rm inf}$. Such a light field during inflation may be realized in models based on the symmetry breaking mechanism: the field can be massless during inflation and it acquires mass at the end of inflation or even after inflation through the symmetry breaking mechanism \cite{Salehian:2020asa,Firouzjahi:2020whk}. However, achieving the nearly massless condition for a spectator field with negligible vev \eqref{X-vev} may not be easy as the stochastic motion generally leads to non-zero $\langle \hat X \rangle$ in our observable universe. The common way to achieve \eqref{X-vev} is to have a potential that makes $X$ settle down to $X=0$, but it requires a large mass which contradicts our assumption. A simple way to overcome this difficulty is to introduce a time-dependent non-minimal coupling of the form $g(t)RX^2$, where $g$ depends on time through the time-dependence of the inflaton, i.e., $g(\phi)=g(\phi(t))$. This time dependence makes the effective mass $m_{\rm eff}^2=m^2+g(t)H^2$ larger than $H^2$ at the early stage when $g={\cal O}(1)$, while $g$ becomes very small $g\ll1$ at the later stage when the scale $k_p$ crosses the horizon. Another possibility is to consider a two-stage inflationary scenario in which $X$ is pinned to $X=0$ due to the large mass in the first stage and becomes almost massless at the second stage. We leave explicit construction of such models, as well as the study of the effects of interactions, for future work.

\vspace{0.7cm}

{\bf Acknowledgments:} MAG thanks Takeshi Kobayashi for valuable discussions. MAG thanks organizers of the workshops ``IBS CTPU-CGA, Tokyo Tech, USTC 2024 summer school and workshop on cosmology, gravity, and particle physics" at Tateyama and ``COSMO'24” at Kyoto University, where this work was presented in its final stages. MAG also thanks Institute of Science Tokyo and Kavli Institute for the Physics and Mathematics of the Universe (IPMU) for hospitality and support during his frequent visits. This work was supported in part by IBS under project code IBS-R018-D3 (MAG) and by JSPS KAKENHI Grant Nos. JP23K03411 (TS), JP20H05853 (MS), and JP24K00624 (MS).
\vspace{0.7cm}

\appendix

\section{Asymptotic form of Kummer's function}\label{app-Kummer}
In this appendix, we present large parameter and large argument behaviour of Kummer's function (confluent hypergeometric of the first kind) $_1F_1\left(a,b;z\right)$ which are useful to find asymptotic expressions for the transfer function in the relativistic and non-relativistic regimes. 

\subsection{Large parameter $a\to\infty$}
When $a\to\infty$ while $b$ and $z$ are fixed, we have\cite{NIST,Temme:1990:UNB}\footnote{One might consider eliminating the factor $i$ in the argument of the Bessel function by working with the modified Bessel function as $I_n(x)=i^{-n} J_n(ix)$. However, the combination $2i a^{\frac{1}{2}} z^{\frac{1}{2}}$ does not necessarily need to be purely imaginary, as both $a$ and $z$ are complex. Indeed, in our case, $2i a^{\frac{1}{2}} z^{\frac{1}{2}}$ turns out to be real to leading order and that is why we preferred to work with the standard Bessel function.}
\begin{align}\label{Hypergeometric-Larg}
_1F_1\left(a,b;z\right)=  e^{\frac{z}{2}-i(1-b)\frac{\pi}{2}} \Gamma\left(b\right)
\frac{z^{\frac{1-b}{2}}}{a^{\frac{1-b}{2}}}
\frac{\Gamma\left(1-a\right)}{\Gamma\left(b-a\right)}
\left(J_{b-1}\left(2i{a^{\frac{1}{2}}}z^{\frac{1}{2}}
\right)\sum_{s=0}^{\infty}\frac{p_{s}(z)}{a^{s}} + i \frac{z^{\frac{1}{2}}}{a^{\frac{1}{2}}}J_{b}\left(2i{a^{\frac{1}{2}}}z^{\frac{1}{2}}\right)\sum_{s=0}^{\infty}\frac{q_{s}(z)}{a^{s}}\right),
\end{align}
where $0<{\rm arg}[a]<\pi$, $-\pi<{\rm arg}[z]\leq\pi$, $J_n(x)$ is the Bessel function of the first kind and
\begin{align}\label{p-def}
p_{k}(z)&=\sum_{s=0}^{k}\genfrac{(}{)}{0.0pt}{}{k}{s}{\left(1-b+s\right)_{k-s}}%
z^{s}c_{k+s}(z), \\
\label{q-def}
q_{k}(z)&=\sum_{s=0}^{k}\genfrac{(}{)}{0.0pt}{}{k}{s}{\left(2-b+s\right)_{k-s}}%
z^{s}c_{k+s+1}(z) \,,
\end{align}
where $(n)_{m} = \Gamma(n+m)/\Gamma(n)$ is the Pochhammer’s symbol, $c_0(z)=1$ and
\begin{align}\label{c-k}
(k+1)c_{k+1}(z)+\sum_{s=0}^{k}\left(\frac{bB_{s+1}}{(s+1)!}+\frac{z(s+1)B_{s+2}}{(s+2)!}\right)c_{k-s}(z)=0;
\qquad
k = 0, 1, 2, \cdots \,,
\end{align}
in which $B_{s}$ are Bernoulli numbers. For example, we find
\begin{align}
\begin{split}
c_1(z) &= \frac{b}{2}-\frac{z}{12} \,,
\qquad
c_2(z) = \frac{1}{24} b (3 b-1) -\frac{b z}{24} + \frac{z^2}{288} \,,
\\
c_3(z) &= \frac{1}{48} (b-1) b^2 + \frac{b z^2}{576} - \frac{(5b (3 b-1) - 2) z}{1440} - \frac{z^3}{10368} \,.
\end{split}
\label{c-explicit}
\end{align}

Substituting the above results in \eqref{p-def} and \eqref{q-def} we find
\begin{align}\label{p-q-explicit}
\begin{split}
p_0(z) &= 1 \,,
\qquad
p_1(z) = - \frac{1}{2} b (b-1) + \frac{1}{24} \left(3 b^2+b-2\right) z-\frac{b z^2}{24} +\frac{z^3}{288}
\qquad
q_0(z) = \frac{b}{2}-\frac{z}{12} \,,
\\
q_1(z) &= -\frac{1}{24} b (b-2) (3 b-1) 
+\frac{1}{48} b \left(b^2+b-4\right) z
+ \frac{1}{480} \left(4-5 b^2\right) z^2
+\frac{b z^3}{576}
-\frac{z^4}{10368} \,.
\end{split}
\end{align}

\subsubsection{$|z|\ll1$}
The so-called Slater's expansion deals with the large parameter $a$ that is written in the form $a=u^2/4+b/2$ such that $u\to\infty$ guaranties $a\to\infty$ \cite{temme2022remarks}. This form is appropriate for our purpose. We consider the following case
\begin{align}\label{Lz-param}
a = \frac{i}{4\mu_k} + \frac{b}{2} \,,
\qquad
z = i \mu \,,
\end{align}
where $\mu_k$ and $\mu$ are real variables and $\mu_k\to0$ guaranties $a\to\infty$. More precisely, large $a$ expansion means that $|a|$ is much larger than $|z|$ and $|b|$. As $b=3/2$ in our case, the large $a$ expansion here coincides with the relativistic limit \eqref{modes-R-mu} as $\mu \mu_k\ll1$. Looking at \eqref{mu-def}, we see that this limit can be achieved for a fixed values of $H$ and $H_k$ and $\m\to0$. However, one needs to be careful that
\begin{align}\label{x-def-app}
x=\sqrt{\mu/\mu_k} \,,
\end{align}
is independent of mass $\m$. In the following, we look at the case when $x$ is fixed while $\mu_k\ll1$ and $\mu\mu_k\ll1$.

Our aim is to substitute \eqref{Lz-param} in \eqref{Hypergeometric-Larg} and then expand it for $\mu_k\ll1$ and $\mu\mu_k\ll1$ up to first order while $x$ is kept fixed. For the ratio of Gamma functions, we find \cite{NIST}
\begin{align}\label{Gamma-expansion}
\frac{\Gamma\left(1-a\right)}{\Gamma\left(b-a\right)} = \frac{\Gamma\big( 1 - \frac{i}{4\mu_k} - \frac{b}{2} \big)}{\Gamma\big(- \frac{i}{4\mu_k} + \frac{b}{2}\big)} = \left(4i\mu_k\right)^{b-1}
\left[ 1 - \frac{2}{3} b (b-1) (b-2)  \mu_k^2 \right] \,.
\end{align}
Now, we assume $|z|\ll1$ or equivalently $\mu\ll1$. Substituting \eqref{Lz-param} and the above result in \eqref{Hypergeometric-Larg}, keeping only the first three terms ($s=0,1,2$) in the summations for $\mu\ll1$, to first order in $\mu\mu_k$ and second orders in $\mu_k$ and $\mu$, we find
\begin{align}\label{Slater-0}
\begin{split}
e^{-\frac{z}{2}}\,_1F_1\left(a,b;z\right) &\approx \Gamma (b)
\bigg[ 
\Big( 1 - \frac{1}{6} (2-b) \mu \mu_k \Big) \frac{J_{b-1}(x)}{(x/2)^{b-1}}
+ \frac{1}{12} \left( 2b(2-b) \mu \mu_k - \mu^2 \right) \frac{J_b(x)}{(x/2)^b}
\bigg] 
\,.
\end{split}
\end{align}
For $b=3/2$, using \eqref{x-def-app} we find
\begin{align}\label{Slater}
e^{-\frac{z}{2}}\,_1F_1\left(a,3/2;z\right) =
\frac{\sin (x)}{x}
- \frac{1}{4} \mu \mu_k
\bigg[
\Big(1-\frac{1}{x^2}\Big) \frac{\sin (x)}{x}
- \left(\frac{2}{3}-\frac{1}{x^2}\right)
\cos (x)
\bigg]
\,.
\end{align}


\subsubsection{$1\ll|z|\ll|a|$}

The large but finite argument limit $|z|\gg1$ is more subtle since we need to be careful about the condition $|z| \ll |a|$. For instance, we cannot directly take $z\to\infty$ limit of Kummer's function. In this case, we still need to deal with \eqref{Hypergeometric-Larg} to make sure that condition $|z| \ll |a|$ satisfies. Thus, for the regime $1\ll|z|\ll|a|$, we look at the terms in \eqref{Hypergeometric-Larg} which have highest power in $z$. 

From \eqref{c-k}, we see that the first term in the summation with $s=0$ is dominant for $|z|\gg1$ such that
\begin{align}\label{c-k-h-z}
c_{k+1}(z)= - \frac{1}{12} \left( \frac{z}{k+1}\right)c_{k}(z) \,,
\end{align}
where we have substituted $B_2=1/6$. From the above relation, we find
\begin{align}\label{c-1-sol}
c_1(z) = - \frac{1}{12} z \,.
\end{align}
It is easy to see that Eq. \eqref{c-k-h-z} has the following solution
\begin{align}\label{c-k-sol}
c_{k}(z) = \frac{c_1(z)^k}{k!} \,.
\end{align}
The highest power of $z$ in \eqref{p-def} and \eqref{q-def} are given by the last term in the summation $s=k$. We thus find
\begin{align}
\begin{split}
p_{k}(z)& \approx z^{k}c_{2k}(z) = \frac{1}{12^{2k}} \frac{z^{3k}}{(2k)!} \,, \\
q_{k}(z)& \approx z^{k}c_{2k+1}(z) = - \frac{1}{12^{2k+1}} \frac{z^{3k+1}}{(2k+1)!} \,,
\end{split}
\label{q-def-h-z}
\end{align}
where we have substituted \eqref{c-k-sol} and \eqref{c-1-sol} in the last steps. One can check that results \eqref{c-k-sol} and \eqref{q-def-h-z} correctly reproduce the highest powers of $z$ in \eqref{c-explicit} and \eqref{p-q-explicit}.

In the above analysis, for any fixed $s$, we have only kept the highest power of $z$ that is $z^{3k}$ and $z^{3k+1}$ in the series of $p_k(z)$ and $q_k(z)$ in \eqref{p-def} and \eqref{q-def} respectively. The next leading term is $z^{3k-1}$ and $z^{3k}$ in $p_k(z)$ and $q_k(z)$ respectively. Following the same strategy as above, we find the next order to the leading order solution \eqref{c-k-sol} as
\begin{align}\label{c-k-sol-next}
c_{k}(z) = \frac{c_1(z)^k}{k!} + \frac{b}{2} \frac{c_1(z)^{k-1}}{(k-1)!} + \cdots \,,
\end{align}
where $c_1(z)$ is given by \eqref{c-1-sol} and $\cdots$ denotes terms which includes $c_1(z)^{n}$ with $n\leq k-2$. 

To keep the next to the leading order terms $z^{3k-1} \subset{p}_k(z)$ and $z^{3k} \subset{q}_k(z)$, we have to fix $s=k$ in \eqref{p-def} as $s=1$ does not contribute. We thus find
\begin{align}
\begin{split}
p_{k}(z)&= \frac{1}{12^{2k}} \frac{z^{3k}}{(2k)!} 
- \frac{b}{2} \frac{1}{12^{2k-1}} \frac{z^{3k-1}}{(2k-1)!} 
+ \cdots \,, \\
q_{k}(z)&= - \frac{1}{12^{2k+1}} \frac{z^{3k+1}}{(2k+1)!} 
+ \frac{b}{2} \frac{1}{12^{2k}} \frac{z^{3k}}{(2k)!} 
+ \cdots\,.
\end{split}
\label{q-def-h-z-f}
\end{align}

Substituting\eqref{q-def-h-z-f} in \eqref{Hypergeometric-Larg} and then performing the summation over $s$, we find
\begin{align}\label{F-R-hz-0}
\begin{split}
e^{-\frac{z}{2}}\,_1F_1\left(a,b;z\right) &=  \Gamma\left(b\right) e^{-i(1-b)\frac{\pi}{2}}
\frac{z^{\frac{1-b}{2}}}{a^{\frac{1-b}{2}}}
\frac{\Gamma\left(1-a\right)}{\Gamma\left(b-a\right)}
\bigg[
J_{b-1}\left(2ia^{\frac{1}{2}}z^{\frac{1}{2}}\right)
\cos \Big(\frac{iz^{3/2}}{12 a^{\frac{1}{2}}}\Big)
- J_{b}\left(2ia^{\frac{1}{2}}z^{\frac{1}{2}}\right) 
\sin \Big(\frac{iz^{3/2}}{12 a^{\frac{1}{2}}}\Big)
\\
&+ i \frac{b}{2} \frac{z^{\frac{1}{2}}}{a^{\frac{1}{2}}} \left( 
J_{b-1}\left(2ia^{\frac{1}{2}}z^{\frac{1}{2}}\right) \sin \Big(\frac{iz^{3/2}}{12a^{\frac{1}{2}}}\Big)
+ J_{b}\left(2ia^{\frac{1}{2}}z^{\frac{1}{2}}\right) \cos \Big(\frac{iz^{3/2}}{12a^{\frac{1}{2}}}\Big)
\right)
+ \cdots  
\bigg] \,,
\end{split}
\end{align}
where the terms $\cdots$ are suppressed by powers of $z/a$.\footnote{The result \eqref{F-R-hz-0} is the solution of \eqref{Transfer-function} in the regime $\frac{\cs{k}}{a}\gg\m\gg{H}$ where the so-called WKB approximation holds as well. The WKB approach gives the solution $T(k,\tau)\propto \frac{1}{a\sqrt{\omega}}\exp{\pm{i}\int\omega\D\tau}$, where $\omega=\sqrt{\cs^2k^2+\m^2{a}^2}= \cs{k} \left(1+\frac{\mu^2}{2x^2} - \frac{\mu^4}{8x^4} + \cdots \right)$ in radiation dominated era with $a=a_i(\tau/\tau_i)$. The physical phase of the solution is given by $\int\omega\D\tau = x \left(1+\frac{\mu^2}{6x^2} - \frac{\mu^4}{40x^4} + \cdots \right)$. To the leading order, this phase corresponds to $2\sqrt{az}-\frac{1}{12}\frac{z^{3/2}}{a^{1/2}}-\frac{1}{320}\frac{z^{5/2}}{a^{3/2}}$. The first two terms can be clearly recovered from \eqref{F-R-hz-0} while it is not clear whether the last term ${\cal O}\left(\frac{z^{5/2}}{a^{3/2}}\right)$ can be recovered or not. This comparison with the WKB approach shows that at least we can trust \eqref{F-R-hz-0} for $1\ll|z|\ll|a|^{3/5}\ll|a|$. This is enough for our purpose in this paper.} 

Keeping the dominant terms in the limit $|z|\ll|a|$ in \eqref{F-R-hz-0}, for $b=3/2$ and \eqref{Lz-param}, we find
\begin{align}\label{F-R-hz}
e^{-\frac{z}{2}}\,_1F_1\left(a,3/2;z\right) = \frac{\sin (x) }{x} \cos \left(\frac{\mu^2}{6 x}\right)
- \frac{1}{x} \left( \frac{\sin(x)}{x} - \cos (x) \right) \sin \left(\frac{\mu ^2}{6 x}\right) \,.
\end{align}

\subsection{Large argument $z\to\infty$}
When $z\to\infty$ and $-\pi/2<\pm{\rm arg}[z]<3\pi/2$, the following expansion holds \cite{NIST}
\begin{align}\label{Kummer-Lz}
_1F_1\left(a,b;z\right) = \frac{\Gamma\left(b\right)e^{z}}{\Gamma\left(a\right)z^{b-a}} 
\sum_{s=0}^{\infty} {\left(1-a\right)_{s}}{\left(b-a\right)_{s}} \frac{z^{-s}}{s!}
+
\frac{\Gamma\left(b\right)e^{\pm\pi\mathrm{i}a}}{\Gamma\left(b-a\right)z^{a}}
\sum_{s=0}^{
	\infty} {\left(a\right)_{s}}{\left(a-b+1\right)_{s}} \frac{(-z)^{-s}}{s!} .
\end{align}
Let us focus on the case \eqref{Lz-param} in which $\mu_k$ and $\mu$ are real variables. Note that we have $\bar{a}=b-a$ for this choice. Since we have assumed that $z$ is pure imaginary, we have ${\rm arg}[z]=\pi/2$ which belongs to the positive branch of \eqref{Kummer-Lz}. Substituting \eqref{Lz-param} in the positive branch of \eqref{Kummer-Lz} we find
\begin{align}
\begin{split}
&e^{-\frac{z}{2}}\,_1F_1\left(a,b;z\right) = 2
{\rm Re}\left[\frac{\Gamma\left(b\right) e^{z/2}}{\Gamma\left(a\right) z^{{\bar a}}}
\sum_{s=0}^{\infty} {\left(1-a\right)_{s}}{\left({\bar a}\right)_{s}} \frac{z^{-s}}{s!}
\right] 
\\
&\hspace{1.5cm} 
= 2 \Gamma\left(b\right) \mu^{-\frac{b}{2}} e^{-\frac{\pi}{8\mu_k}}
{\rm Re}\left[
\frac{e^{\frac{i\mu}{2} - \frac{i\pi{b}}{4}} \mu^{\frac{i}{4\mu_k}}}{\Gamma\big(\frac{b}{2}+\frac{i}{4\mu_k} \big)}
\sum_{s=0}^{\infty} \left(1-\frac{b}{2}-\frac{i}{4\mu_k}\right)_{s} \left(\frac{b}{2}-\frac{i}{4\mu_k}\right)_{s} \frac{\left(i\mu\right)^{-s}}{s!}
\right] 
.
\end{split}
\label{Kummer-Lz-p}
\end{align}

\section{Power spectrum of dark matter density contrast}\label{app-PS-DM}

As it can be seen from the total energy density \eqref{rho-x}, there are inhomogeneous fluctuations around the homogeneous vacuum average value \eqref{rho}. In this appendix, we study the behevior of these fluctuations.

When the mass term dominates after inflation ($f=1$), the energy density of the dark matter is given by
\begin{align}
\rho_X(\tau,x) \approx \frac{1}{2} m^2 X^2 \, 
\Rightarrow 
 {\bar \rho}_X(\tau) \approx \frac{1}{2} m^2 \langle X^2 \rangle \,,
\end{align}
where 
\begin{align}\label{X-v}
 \langle X^2 \rangle = \int\D\ln{k}\, {\cal P}_X(k,\tau) \,.
\end{align}
The dark matter density contrast \eqref{delta} will be
\begin{align}
\delta_{X,{\bf k}}(\tau) \equiv \frac{1}{\langle{X}^2\rangle} \int \frac{\D^3q}{(2\pi)^3} X_{\bf{k-q}}(\tau) X_{\bf q}(\tau) \,.
\end{align}
The two-point function is
\begin{align}\label{delta-2PF}
\langle \delta_{X,{\bf k}}(\tau) \delta_{X,{\bf{k}'}}(\tau)\rangle
&= 
\frac{1}{\langle{X}^2\rangle^2} \int \frac{\D^3q}{(2\pi)^3} \frac{\D^3q'}{(2\pi)^3}
\Big[
\langle X_{\bf{k-q}}(\tau) X_{\bf{k}'-\bf{q}'}(\tau) \rangle 
\langle X_{\bf{q}}(\tau) X_{\bf{q}'}(\tau) \rangle 
\\
&+
\langle X_{\bf{k-q}}(\tau) X_{\bf{q}'}(\tau) \rangle 
\langle X_{\bf{q}}(\tau) X_{\bf{k}'-\bf{q}'}(\tau) \rangle 
\Big] 
\\
&= (2\pi)^3 \delta({\bf k}+{\bf k}') \frac{8\pi^4}{\langle{X}^2\rangle^2} 
\int \frac{\D^3q}{(2\pi)^3} \frac{{\cal P}_X(|{\bf k}-{\bf q}|,\tau){\cal P}_X(q,\tau)}{|{\bf k}-{\bf q}|^3q^3}
\,.
\end{align}
Using the definition of power spectrum, from the above result one can show that
\begin{align}\label{delta-V-app}
\langle{\delta_X^2}\rangle = \int \D\ln{k}\, {\cal P}_{\delta_X}(k,\tau) = 2 \,,
\end{align}
where we have changed the integration measure as ${\bf k}\to{\bf k}+{\bf q}$ and then we have used \eqref{X-v}. The above result coincides with \eqref{delta-V} which is obtained through the Gaussianity of $X$.

In order to perform direct calculations for the case of sharply peaked power spectrum, it is better to rewrite the expression for the power spectrum of the density contrast, which is defined by \eqref{delta-2PF}, as follows
\begin{align}\label{PS-delta-general}
{\cal P}_{\delta_X}(k,\tau) = \frac{k^2}{\langle{X}^2\rangle^2} \int_0^\infty \frac{\D{q}}{q^2} {\cal P}_X(q,\tau) \int_{|k-q|}^{|k+q|} \frac{\D{p}}{p^2} {\cal P}_X(p,\tau) \,,
\end{align}
where we have defined $p\equiv|{\bf k}-{\bf q}|$. Substituting the sharply peaked power spectrum \eqref{PS-delta}, we find
\begin{align}\label{P-delta-SP}
{\cal P}_{\delta_X}(k,k_p) = \left(\frac{k}{k_p}\right)^2 \Theta (2k_p-k) \,.
\end{align}
The power spectrum is completely suppressed at scales $k \ll k_p$ while it becomes of the order of unity at $k=k_p$. Substituting \eqref{P-delta-SP} in \eqref{delta-V-app} one can directly confirm the general result \eqref{delta-V-app} for this particular case. Thus, there are ${\cal O}(1)$ fluctuations around the peak $k=k_p$. 

\bibliographystyle{JHEPmod}
\bibliography{refs}

\end{document}